\pdfoutput=1
\documentclass[12pt,a4paper]{article}

\usepackage{ifthen} 
\usepackage{enumerate}
\newboolean{pdflatex}
\setboolean{pdflatex}{true} 

\newboolean{articletitles}
\setboolean{articletitles}{true} 

\newboolean{uprightparticles}
\setboolean{uprightparticles}{false} 

\def\paperauthors{LHCb collaboration} 
\def\paperasciititle{LFUV hadronic 13 TeV} 
\def\papertitle{Test of lepton flavor universality using $B^0 \to D^{*-}\tau^+\nu_{\tau}$ decays with hadronic $\tau$ channels }
\def\paperkeywords{{High Energy Physics}, {LHCb}} 
\def\papercopyright{\the\year\ CERN for the benefit of the LHCb collaboration} 
\def\paperlicence{CC BY 4.0 licence}
\def\paperlicenceurl{https://creativecommons.org/licenses/by/4.0/}

\usepackage{bm}
\usepackage{siunitx}
\usepackage{collcell}

\usepackage{longtable} 

\usepackage[top=1in, bottom=1.25in, left=1in, right=1in]{geometry}

\usepackage{microtype}
\usepackage{lineno}  
\usepackage{xspace} 
\usepackage{caption} 

\usepackage{graphicx}  
\usepackage{color}
\usepackage{colortbl}
\graphicspath{{./figs/},{./eps/}} 

\usepackage{amsmath} 
\usepackage{amssymb}
\usepackage{amsfonts}
\usepackage{upgreek} 

\newcommand*\patchAmsMathEnvironmentForLineno[1]{%
\expandafter\let\csname old#1\expandafter\endcsname\csname #1\endcsname
\expandafter\let\csname oldend#1\expandafter\endcsname\csname
end#1\endcsname
 \renewenvironment{#1}%
   {\linenomath\csname old#1\endcsname}%
   {\csname oldend#1\endcsname\endlinenomath}%
}
\newcommand*\patchBothAmsMathEnvironmentsForLineno[1]{%
  \patchAmsMathEnvironmentForLineno{#1}%
  \patchAmsMathEnvironmentForLineno{#1*}%
}
\AtBeginDocument{%
\patchBothAmsMathEnvironmentsForLineno{equation}%
\patchBothAmsMathEnvironmentsForLineno{align}%
\patchBothAmsMathEnvironmentsForLineno{flalign}%
\patchBothAmsMathEnvironmentsForLineno{alignat}%
\patchBothAmsMathEnvironmentsForLineno{gather}%
\patchBothAmsMathEnvironmentsForLineno{multline}%
\patchBothAmsMathEnvironmentsForLineno{eqnarray}%
}

\usepackage{hyperxmp}

\usepackage[pdftex,
            pdfauthor={\paperauthors},
            pdftitle={\paperasciititle},
            pdfkeywords={\paperkeywords},
            pdfcopyright={Copyright (C) \papercopyright},
            pdflicenseurl={\paperlicenceurl}]{hyperref}

\usepackage[colorinlistoftodos,textsize=scriptsize]{todonotes}

\usepackage[bottom,flushmargin,hang,multiple]{footmisc}

\usepackage[all]{hypcap} 

\usepackage{xspace} 
\usepackage{upgreek}

\def\lhcb   {\mbox{LHCb}\xspace}

\def\babar  {\mbox{BaBar}\xspace}

\def\MagUp {\mbox{\em Mag\kern -0.05em Up}\xspace}

\ifthenelse{\boolean{uprightparticles}}%
{

 \def\Peta        {\ensuremath{\upeta}\xspace}

 \def\Pmu         {\ensuremath{\upmu}\xspace}                 
 \def\Pnu         {\ensuremath{\upnu}\xspace}                 
                  
 \def\Ppi         {\ensuremath{\uppi}\xspace}                 
                  
 \def\Prho        {\ensuremath{\uprho}\xspace}                 
                  
 \def\Ptau        {\ensuremath{\uptau}\xspace}

 \def\PDelta      {\ensuremath{\Delta}\xspace}                 
 \def\PXi         {\ensuremath{\Xi}\xspace}                 
 \def\PLambda     {\ensuremath{\Lambda}\xspace}                 
 \def\PSigma      {\ensuremath{\Sigma}\xspace}                 
 \def\POmega      {\ensuremath{\Omega}\xspace}                 
 \def\PUpsilon    {\ensuremath{\Upsilon}\xspace}
 \let\oldPi\Pi
 \def\PPi         {\ensuremath{\oldPi}\xspace}

 \def\PB      {\ensuremath{\mathrm{B}}\xspace}                 
                  
 \def\PD      {\ensuremath{\mathrm{D}}\xspace}

 \def\PK      {\ensuremath{\mathrm{K}}\xspace}

 \def\PX      {\ensuremath{\mathrm{X}}\xspace}

 \def\Pb      {\ensuremath{\mathrm{b}}\xspace}                 
 \def\Pc      {\ensuremath{\mathrm{c}}\xspace}

 \def\Pi      {\ensuremath{\mathrm{i}}\xspace}

 \def\Ps      {\ensuremath{\mathrm{s}}\xspace}

 \def\thebaroffset{0.0em}
}
{

 \def\Peta        {\ensuremath{\eta}\xspace}

 \def\Pmu         {\ensuremath{\mu}\xspace}                 
 \def\Pnu         {\ensuremath{\nu}\xspace}                 
                  
 \def\Ppi         {\ensuremath{\pi}\xspace}                 
                  
 \def\Prho        {\ensuremath{\rho}\xspace}                 
                  
 \def\Ptau        {\ensuremath{\tau}\xspace}

 \mathchardef\PDelta="7101
 \mathchardef\PXi="7104
 \mathchardef\PLambda="7103
 \mathchardef\PSigma="7106
 \mathchardef\POmega="710A
 \mathchardef\PUpsilon="7107
 \mathchardef\PPi="7105
                  
 \def\PB      {\ensuremath{B}\xspace}                 
                  
 \def\PD      {\ensuremath{D}\xspace}

 \def\PK      {\ensuremath{K}\xspace}

 \def\PX      {\ensuremath{X}\xspace}

 \def\Pb      {\ensuremath{b}\xspace}                 
 \def\Pc      {\ensuremath{c}\xspace}

 \def\Pi      {\ensuremath{i}\xspace}

 \def\Ps      {\ensuremath{s}\xspace}

 \def\thebaroffset{0.18em}
}
\newcommand{\offsetoverline}[2][\thebaroffset]{\kern #1\overline{\kern -#1 #2}}%

\makeatletter
\ifcase \@ptsize \relax
  \newcommand{\miniscule}{\@setfontsize\miniscule{4}{5}}
\or
  \newcommand{\miniscule}{\@setfontsize\miniscule{5}{6}}
\or
  \newcommand{\miniscule}{\@setfontsize\miniscule{5}{6}}
\fi
\makeatother

\DeclareRobustCommand{\optbar}[1]{\shortstack{{\miniscule (\rule[.5ex]{1.25em}{.18mm})}
  \\ [-.7ex] $#1$}}

\def\mup        {{\ensuremath{\Pmu^+}}\xspace}

\def\tauon      {{\ensuremath{\Ptau}}\xspace}
\def\taup       {{\ensuremath{\Ptau^+}}\xspace}

\def\neu        {{\ensuremath{\Pnu}}\xspace}
\def\neub       {{\ensuremath{\overline{\Pnu}}}\xspace}
\def\neum       {{\ensuremath{\neu_\mu}}\xspace}

\def\neut       {{\ensuremath{\neu_\tau}}\xspace}
\def\neutb      {{\ensuremath{\neub_\tau}}\xspace}

\def\squark    {{\ensuremath{\Ps}}\xspace}

\def\cquark    {{\ensuremath{\Pc}}\xspace}

\def\bquark    {{\ensuremath{\Pb}}\xspace}
\def\bquarkbar {{\ensuremath{\overline \bquark}}\xspace}
\def\bbbar     {{\ensuremath{\bquark\bquarkbar}}\xspace}

\def\pion   {{\ensuremath{\Ppi}}\xspace}
\def\piz    {{\ensuremath{\pion^0}}\xspace}
\def\pip    {{\ensuremath{\pion^+}}\xspace}
\def\pim    {{\ensuremath{\pion^-}}\xspace}

\def\rhomeson {{\ensuremath{\Prho}}\xspace}

\def\rhop     {{\ensuremath{\rhomeson^+}}\xspace}

\def\kaon    {{\ensuremath{\PK}}\xspace}

\def\KorKbar {\kern \thebaroffset\optbar{\kern -\thebaroffset \PK}{}\xspace}
\def\Kz      {{\ensuremath{\kaon^0}}\xspace}

\def\Kp      {{\ensuremath{\kaon^+}}\xspace}
\def\Km      {{\ensuremath{\kaon^-}}\xspace}

\newcommand{\etapr}{\ensuremath{\Peta^{\prime}}\xspace}

\def\Dbar    {{\ensuremath{\offsetoverline{\PD}}}\xspace}
\def\D       {{\ensuremath{\PD}}\xspace}
\def\Db      {{\ensuremath{\Dbar}}\xspace}
\def\DorDbar {\kern \thebaroffset\optbar{\kern -\thebaroffset \PD}\xspace}
\def\Dz      {{\ensuremath{\D^0}}\xspace}
\def\Dzb     {{\ensuremath{\Dbar{}^0}}\xspace}
\def\Dp      {{\ensuremath{\D^+}}\xspace}
\def\Dm      {{\ensuremath{\D^-}}\xspace}

\def\DpDm    {\ensuremath{\Dp {\kern -0.16em \Dm}}\xspace}
\def\Dstar   {{\ensuremath{\D^*}}\xspace}

\def\Dstarp  {{\ensuremath{\D^{*+}}}\xspace}
\def\Dstarm  {{\ensuremath{\D^{*-}}}\xspace}

\def\Ds      {{\ensuremath{\D^+_\squark}}\xspace}
\def\Dsp     {{\ensuremath{\D^+_\squark}}\xspace}

\def\Dss     {{\ensuremath{\D^{*+}_\squark}}\xspace}

\def\B       {{\ensuremath{\PB}}\xspace}

\def\BorBbar {\kern \thebaroffset\optbar{\kern -\thebaroffset \PB}\xspace}
\def\Bz      {{\ensuremath{\B^0}}\xspace}

\def\Bd      {{\ensuremath{\B^0}}\xspace}

\def\BdorBdbar {\kern \thebaroffset\optbar{\kern -\thebaroffset \Bd}\xspace}

\def\Bs      {{\ensuremath{\B^0_\squark}}\xspace}

\def\BsorBsbar {\kern \thebaroffset\optbar{\kern -\thebaroffset \Bs}\xspace}

\def\Y#1S{\ensuremath{\PUpsilon{(#1S)}}\xspace}

\def\LorLbar     {\kern \thebaroffset\optbar{\kern -\thebaroffset \PLambda}\xspace}

\def\BF         {{\ensuremath{\mathcal{B}}}\xspace}

\newcommand{\decay}[2]{\ensuremath{#1\!\to #2}\xspace} 

\def\to                 {\ensuremath{\rightarrow}\xspace}

\def\AT#1     {\ensuremath{A_{\mathrm{T}}^{#1}}\xspace}           

\def\C#1      {\ensuremath{\mathcal{C}_{#1}}\xspace}                       
\def\Cp#1     {\ensuremath{\mathcal{C}_{#1}^{'}}\xspace}                    
\def\Ceff#1   {\ensuremath{\mathcal{C}_{#1}^{\mathrm{(eff)}}}\xspace}        
\def\Cpeff#1  {\ensuremath{\mathcal{C}_{#1}^{'\mathrm{(eff)}}}\xspace}       
\def\Ope#1    {\ensuremath{\mathcal{O}_{#1}}\xspace}                       
\def\Opep#1   {\ensuremath{\mathcal{O}_{#1}^{'}}\xspace}                    

\newcommand{\nospaceunit}[1]{\ensuremath{\text{#1}}}       
\newcommand{\aunit}[1]{\ensuremath{\text{\,#1}}}       

\newcommand{\tev}{\aunit{Te\kern -0.1em V}\xspace}
\newcommand{\gev}{\aunit{Ge\kern -0.1em V}\xspace}
\newcommand{\mev}{\aunit{Me\kern -0.1em V}\xspace}
\newcommand{\kev}{\aunit{ke\kern -0.1em V}\xspace}
\newcommand{\ev}{\aunit{e\kern -0.1em V}\xspace}
 
\newcommand{\mevc}{\ensuremath{\aunit{Me\kern -0.1em V\!/}c}\xspace}
\newcommand{\gevc}{\ensuremath{\aunit{Ge\kern -0.1em V\!/}c}\xspace}
\newcommand{\mevcc}{\ensuremath{\aunit{Me\kern -0.1em V\!/}c^2}\xspace}
\newcommand{\gevcc}{\ensuremath{\aunit{Ge\kern -0.1em V\!/}c^2}\xspace}

\def\mm   {\aunit{mm}\xspace}

\def\mum  {\ensuremath{\,\upmu\nospaceunit{m}}\xspace}

\def\fb   {\ensuremath{\aunit{fb}}\xspace}
\def\invfb   {\ensuremath{\fb^{-1}}\xspace}

\newcommand{\stat}{\aunit{(stat)}\xspace}
\newcommand{\syst}{\aunit{(syst)}\xspace}

\def\gsim{{~\raise.15em\hbox{$>$}\kern-.85em
          \lower.35em\hbox{$\sim$}~}\xspace}
\def\lsim{{~\raise.15em\hbox{$<$}\kern-.85em
          \lower.35em\hbox{$\sim$}~}\xspace}

\def\pt         {\ensuremath{p_{\mathrm{T}}}\xspace}

\def\ptot       {\ensuremath{p}\xspace}

\def\evtgen     {\mbox{\textsc{EvtGen}}\xspace}

\def\geant      {\mbox{\textsc{Geant4}}\xspace}

\def\photos     {\mbox{\textsc{Photos}}\xspace}

\def\pythia     {\mbox{\textsc{Pythia}}\xspace}

\def\tell1  {TELL1\xspace}
\def\ukl1   {UKL1\xspace}

\newcommand{\eg}{\mbox{\itshape e.g.}\xspace}
\newcommand{\ie}{\mbox{\itshape i.e.}\xspace}

\newcommand{\lhcborcid}[1]{\href{https://orcid.org/#1}{\hspace*{0.1em}\raisebox{-0.45ex}{\includegraphics[width=1em]{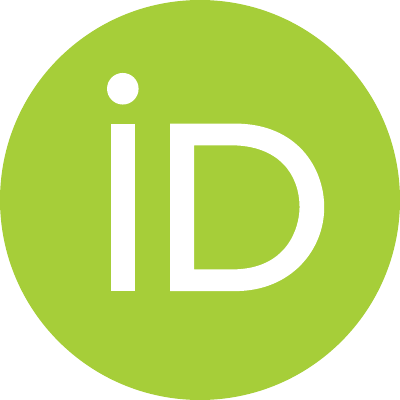}}}}

\usepackage{cite} 
\usepackage{mciteplus}

\usepackage{xspace} 
\usepackage{upgreek}

\def\rdstar {\ensuremath{\mathcal{R}(\Dstar)}\xspace}
\def\rdstarm {\ensuremath{\mathcal{R}(\Dstarm)}\xspace}
\def\rd {\ensuremath{\mathcal{R}(D)}\xspace}

\def\sig {\ensuremath{\decay{\Bd}{\Dstarm\taup\neut}}\xspace}
\def\signalmode {\mbox{\ensuremath{\decay{\Bd}{\Dstarm\taup\neut}}}\xspace}
\def\norm {\mbox{\ensuremath{\decay{\Bd}{\Dstarm3\pi}}}\xspace}
\def\normfull {\mbox{\ensuremath{\decay{\Bd}{\Dstarm\pip\pim\pip}}}\xspace}

\newcommand{\tevcc}{\ensuremath{\aunit{Te\kern -0.1em V\!/}c^2}\xspace}

\def\kappadstar {\ensuremath{\mathcal{K}(\Dstarm)}\xspace}

\def\tauola     {\mbox{\textsc{Tauola}}\xspace}

\def\Dssz{\ensuremath{\D_{\squark 0}^{*+}}\xspace}
\def\Dsprime{\ensuremath{\D_{\squark 1}^{+}}\xspace}

\def\KDstar{\ensuremath{\mathcal{K}(\Dstarm)}\xspace}
\def\signalmode{\decay{\Bd}{\Dstarm \taup \neut}}

\def\feeddown{\decay{\PB}{\Db^{**} \taup \neut}}

\def\tauthreepi{\decay{\taup}{3\pi \neutb}}
\def\tauthreepipiz{\decay{\taup}{3\pi \piz \neutb}}

\usepackage{cleveref}
\crefname{section}{Sec.}{Secs.}
\crefname{table}{Table}{Tables}
\crefname{figure}{Fig.}{Figs.}

\begin{document}

\renewcommand{\thefootnote}{\fnsymbol{footnote}}
\setcounter{footnote}{1}


\begin{titlepage}
\pagenumbering{roman}

\vspace*{-1.5cm}
\centerline{\large EUROPEAN ORGANIZATION FOR NUCLEAR RESEARCH (CERN)}
\vspace*{1.5cm}
\noindent
\begin{tabular*}{\linewidth}{lc@{\extracolsep{\fill}}r@{\extracolsep{0pt}}}
\ifthenelse{\boolean{pdflatex}}
{\vspace*{-1.5cm}\mbox{\!\!\!\includegraphics[width=.14\textwidth]{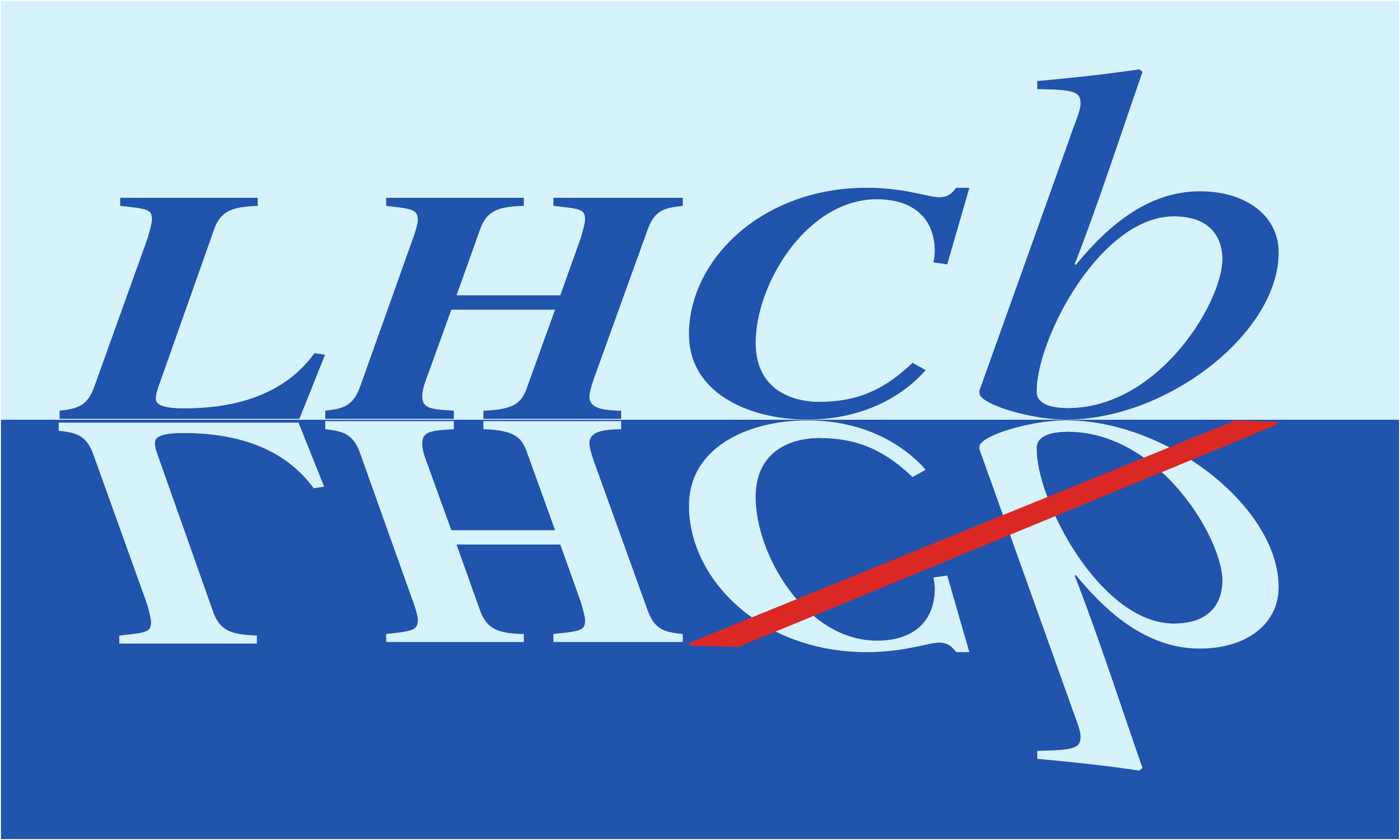}} & &}%
{\vspace*{-1.2cm}\mbox{\!\!\!\includegraphics[width=.12\textwidth]{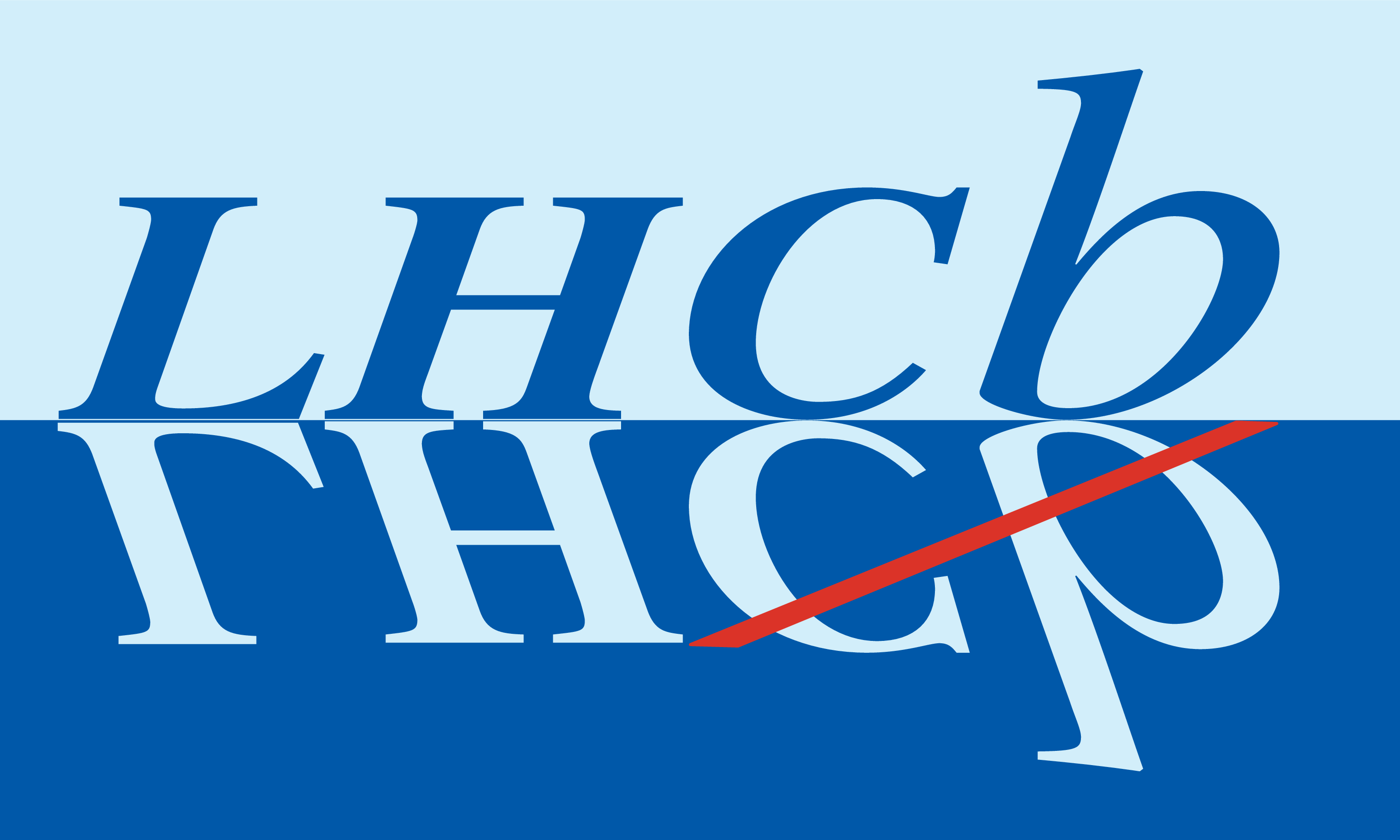}} & &}%
\\
 & & CERN-EP-2023-062 \\  
 & & LHCb-PAPER-2022-052 \\  
 & & 13 May 2024 \\ 
 & & \\
\end{tabular*}

\vspace*{4.0cm}

{\normalfont\bfseries\boldmath\huge
\begin{center}
  \papertitle 
\end{center}
}

\vspace*{1.0cm}

\begin{center}
\paperauthors\footnote{Full author list given at the end of the paper.}
\end{center}

\vspace{\fill}

\begin{abstract}
  \noindent
The branching fraction \mbox{$\BF(\decay{\Bd}{\Dstarm\taup\neut})$} is measured relative to that of the normalization mode \normfull using hadronic \mbox{$\decay{\taup}{\pip\pim\pip(\piz)\neutb}$} decays in proton-proton collision data at a center-of-mass energy of $13\tev$ collected by the \lhcb experiment, corresponding to an integrated luminosity of $2\invfb$. The measured ratio is \mbox{$\BF(\sig)/\BF(\normfull)= 1.79 \pm 0.11 \pm 0.11$}, where the first uncertainty is statistical and the second is related to systematic effects. Using established branching fractions for the \normfull and \mbox{$\decay{\Bd}{D^{*-} \mu^+\nu_\mu}$} modes, the lepton universality test, \mbox{$\rdstarm \equiv \BF(\decay{\Bd}{\Dstarm\taup\neut})/\BF(\decay{\Bd}{\Dstarm\mup\neum})$} is calculated,
$$
\rdstarm =  0.260 \pm 0.015 \pm 0.016 \pm 0.012\, ,
$$
where the third uncertainty is due to the uncertainties on the external branching fractions. This result is consistent with the Standard Model prediction and with previous measurements.

\end{abstract}


\begin{center}
  Published in
  \href{https://link.aps.org/doi/10.1103/PhysRevD.108.012018}{Phys.~Rev.~D 108 (2023) 1, 012018}
\end{center}

\vspace{\fill}

{\footnotesize 
\centerline{\copyright~\papercopyright. \href{\paperlicenceurl}{\paperlicence}.}}
\vspace*{2mm}

\end{titlepage}


\newpage
\setcounter{page}{2}
\mbox{~}
%
%
%
%


\renewcommand{\thefootnote}{\arabic{footnote}}
\setcounter{footnote}{0}

\cleardoublepage


\pagestyle{plain} 
\setcounter{page}{1}
\pagenumbering{arabic}


\section{INTRODUCTION}
\label{sec:Introduction}
Measurements of \mbox{${\cal R}(D^{(*)}) \equiv \BF(\decay{\Bd}{D^{(*)}\taup\neut})/\BF(\decay{\Bd}{D^{(*)}\ell^+\nu_\ell})$}, the ratio of branching fractions with $\ell=\mu,e$, test lepton flavor universality in $b\rightarrow c\ell\nu_{\ell}$ transitions. First measured by the \babar Collaboration in 2012~\cite{Lees_2012}, this ratio has been studied by the Belle~\cite{belle_collaboration_measurement_2015,the_belle_collaboration_measurement_2018} and LHCb~\cite{LHCb-PAPER-2017-017,LHCb-PAPER-2017-027} experiments using hadronic and muonic decay modes of the \taup lepton\footnote{\babar and Belle used decays with both muons and electrons in the ${\cal R}(D^{(*)})$ denominator, while LHCb has exclusively studied decays with muons so far.}. After  the latest LHCb Collaboration result using muonic \taup decays~\cite{LHCb-PAPER-2022-039}, the discrepancy between the world-average values of \rdstar and \rd measurements with their theoretical prediction is at the level of 3 standard deviations\cite{hflav_2021}.  The predicted value by the Standard Model~(SM) of particle physics is $R(D^*)=0.254\pm0.005$~\cite{hflav_2021}. Several extensions to the SM can explain this anomaly, \eg, leptoquark models~\cite{Fajfer_2012,Fajfer_2016,crivellin_simultaneous_2017}, which typically assume a  leptoquark that preferentially couples to third-generation leptons and has a mass below $1\tevcc$.

The LHCb hadronic $\rdstarm$ analysis was first performed using proton-proton~($pp$) collision data collected at center-of-mass energies of $\sqrt{s}=7$ and 8\tev in 2011 and 2012~\cite{LHCb-PAPER-2017-017,LHCb-PAPER-2017-027}, corresponding to an integrated luminosity of $3\invfb$. This paper presents a similar measurement of $\rdstarm$ based on $pp$ collision data taken at 13\tev in 2015 and 2016, corresponding to an integrated luminosity of $2\invfb$. Despite the lower integrated luminosity, the increase of the \bbbar production cross section with the center-of-mass energy by nearly a factor of 2 and improvements in the LHCb trigger provide about 40\% more signal candidates than in the previous analysis. 

The analysis strategy in this paper is similar to that detailed in the previous study~\cite{LHCb-PAPER-2017-017,LHCb-PAPER-2017-027}, and includes changes that improve the signal efficiency. The \taup lepton is reconstructed in the hadronic final state $3\pi(\piz)\neutb$, where \mbox{$3\pi \equiv \pip\pim\pip$}, and the \Dstarm candidate is reconstructed through the $\mbox{\decay{\Dstarm}{\pim\Dzb(\to \Kp\pim)}}$ decay.\footnote{The inclusion of charge-conjugate decay modes is implied throughout the paper.} The \norm decay is chosen as the normalization mode because it has the same visible final state as the signal mode. Many of the systematic uncertainties due to detector and reconstruction effects cancel in the ratio of their branching fractions, defined as
\begin{equation}{\cal{K}}(\Dstar^-)\equiv\frac{{\cal{B}}(B^0 \to D^{*-} \tau^+
    \nu_{\tau})
}{{\cal{B}}(B^0 \to
    D^{*-} 3\pion)} =
    \frac{N_{\mathrm{sig}}}{N_{\mathrm{norm}}}\frac{\varepsilon_{\mathrm{norm}}}{\varepsilon_{\mathrm{sig}}}\frac{1}{{\cal{B}}(\tau^+\to
3\pi\neutb)+{\cal{B}}(\tau^+\to
3\pi\piz\neutb)}\,.
\label{eqn:kappa}
\end{equation}
Here, $N_{\rm sig}$ and $N_{\rm norm}$ are the yields in the signal and normalization modes, respectively, which are obtained from the data. The efficiencies $\varepsilon_{\mathrm{sig}}$ and $\varepsilon_{\mathrm{norm}}$, for the signal and normalization modes, respectively, are determined from the simulation. The signal efficiency $\varepsilon_{\mathrm{sig}}$ is calculated as the average between the \mbox{$\tau^+\to 3\pi\bar{\nu}_{\tau}$} 
and the \mbox{$\tau^+\to 3\pi\pi^0\bar{\nu}_\tau$} decays weighted by their relative branching fractions~\cite{PDG2022}. Finally, using the known branching fractions $\mbox{\BF(\decay{\Bd}{\Dstarm 3\pi})}$ and $\mbox{\BF(\decay{\Bd}{\Dstarm \mup \neum})}$~\cite{PDG2022}, \rdstarm is obtained as
\begin{equation}
 \rdstarm = \mathcal{K}(\Dstarm) \frac{\BF(\decay{\Bd}{\Dstarm 3\pi})}{\BF(\decay{\Bd}{\Dstarm \mup \neum})}\,.
\end{equation}
To avoid biases, the numerical results of this analysis were not examined until the full procedure had been finalized.

This paper is organized as follows. The LHCb detector and simulation are described in \cref{sec:Detector}.  Details of the selection criteria used to select the \sig and \norm candidates are presented in \cref{sec:selection}, followed by \cref{sec:doublecharm}, which describes the study of the double-charm decays of the $B$ meson that form the dominant background for the signal mode. 
The determination of the signal and normalization yields are provided in \cref{sec:yield}, and the systematic uncertainties are discussed in \cref{sec:systematics}.
Finally, the results are given in \cref{sec:conclusion}.

\section{DETECTOR AND SIMULATION}
\label{sec:Detector}
The \lhcb detector~\cite{LHCb-DP-2008-001,LHCb-DP-2014-002} is a single-arm forward
spectrometer covering the pseudorapidity range $2<\eta <5$,
designed for the study of particles containing \bquark or \cquark
quarks. The detector includes a high-precision tracking system
consisting of a silicon-strip vertex detector surrounding the $pp$
interaction region~\cite{LHCb-DP-2014-001}, a large-area silicon-strip detector located
upstream of a dipole magnet with a bending power of about
$4{\mathrm{\,Tm}}$, and three stations of silicon-strip detectors and straw
drift tubes~\cite{LHCb-DP-2013-003,LHCb-DP-2017-001} placed downstream of the magnet.
The tracking system provides a measurement of the momentum, \ptot, of charged particles with
a relative uncertainty that varies from 0.5\% at low momentum to 1.0\% at 200\gevc.
The minimum distance of a track to a primary $pp$ collision vertex~(PV), the impact parameter~(IP), 
is measured with a resolution of $(15+29/\pt)\mum$,
where \pt is the component of the momentum transverse to the beam, in\,\gevc.
Different types of charged hadrons are distinguished using information
from two ring-imaging Cherenkov detectors~\cite{LHCb-DP-2012-003}. 
Photons, electrons and hadrons are identified by a calorimeter system consisting of
scintillating-pad and preshower detectors, an electromagnetic
and a hadronic calorimeter. Muons are identified by a
system composed of alternating layers of iron and multiwire
proportional chambers~\cite{LHCb-DP-2012-002}.

The online event selection is performed by a trigger~\cite{LHCb-DP-2012-004}, 
which consists of a hardware stage, based on information from the calorimeter and muon systems, followed by a software stage, which applies a full event reconstruction.
At the hardware-trigger stage, events are required to have a muon with high \pt or a
hadron, photon or electron with high transverse energy in the calorimeters. 
The hadron can originate from either the decay chain under consideration or the remainder of the event.
The software trigger requires a two-, three- or four-track
secondary vertex with a significant displacement from any PV. At least one charged particle
must have a transverse momentum $\pt > 1.6\gevc$ and be
inconsistent with originating from any PV.
A multivariate algorithm~\cite{BBDT,LHCb-PROC-2015-018} is used for
the identification of secondary vertices consistent with the decay
of a \bquark hadron.

Simulation is required to model the effects of the detector acceptance and the imposed selection requirements.
In the simulation, $pp$ collisions are generated using
\pythia~\cite{Sjostrand:2007gs,*Sjostrand:2006za} 
with a specific \lhcb configuration~\cite{LHCb-PROC-2010-056}.
Decays of unstable particles
are described by \evtgen~\cite{Lange:2001uf}, in which final-state
radiation is generated using \photos~\cite{davidson2015photos}.  The \taup decays to $3\pi\neutb$ and $3\pi\pi^0\neutb$ are simulated using the resonance chiral Lagrangian model~\cite{Chiral_lagrangian} as implemented in the \tauola~\cite{DAVIDSON2012821} package  tuned according to the results from the \babar Collaboration~\cite{NUGENT201438}.
The interaction of the generated particles with the detector, and its response,
are implemented using the \geant
toolkit~\cite{Allison:2006ve, *Agostinelli:2002hh} as described in
Ref.~\cite{LHCb-PROC-2011-006}. 
Large samples of simulated events are required to reduce the systematic uncertainty due to the sample size. A fast simulation technique, \textsc{redecay}~\cite{LHCb-DP-2018-004},  is used for this purpose in which the underlying $pp$ interaction is reused multiple times, with an independently generated signal decay for each. These samples have been validated against simulated events using unique underlying interactions.

\section{EVENT SELECTION}
\label{sec:selection}
The reconstruction of the decay kinematics follows the procedure given in Refs.~\cite{LHCb-PAPER-2017-017,LHCb-PAPER-2017-027}. The signal candidates are formed by combining a \Dstarm meson that decays as \mbox{\decay{\Dstarm}{\pim\Dzb(\to \Kp \pim)}} and a 3$\pi$ system which is detached from the \Bd decay vertex due to the non-negligible lifetime of the \taup lepton. 
The dominant background contribution is \Bd decays in which the $3\pi$ system comes promptly from the \Bd vertex, called prompt background hereafter. To suppress this background, the distance between the \Bd and $3\pi$ vertices along the beam direction, $\Delta z \equiv z(3\pi) - z(\Bd)$, is required to be at least twice its uncertainty~($\sigma_{\Delta z}$). Double-charm $\mbox{\decay{B}{\Dstarm \D (X)}}$ decays\footnote{Throughout the paper, $X$ denotes unreconstructed particles that are known to be present in the decay chain, and $(X)$ stands for those that may or may not be present.} are the next dominant background, with a detached-vertex topology similar to that of the signal decays.
The remaining sources of background are suppressed by requiring the $3\pi$ system to be consistent with originating from a common vertex.

Three categories of data and simulation are used in this analysis: signal, normalization and control. The control samples are used to study the double-charm background and are selected to enrich the number of \Dsp decays. 
The selection process for each set is split into two stages. 
First, common selection criteria are applied to suppress the candidates originating from random combinations of final-state particles for the signal, normalization and control samples.
Second, specific requirements are placed on each set. The selection of control samples is described in \cref{sec:doublecharm}.

\subsection{Common selection criteria}
The purpose of these common requirements is to suppress the prompt and combinatorial backgrounds.
Tracks consistent with a kaon or pion hypothesis and having \mbox{$p>2\gevc$} and \mbox{$\pt>250\mevc$} are 
selected to form \Dzb candidates, which are required to have a mass in the range $[1840,\,1890]\mevcc$ and \pt larger than $1.2\gevc$. The \Dzb candidates are then combined with tracks consistent with the pion hypothesis and with \mbox{$\pt>110\mevc$} to form \Dstarm candidates, where the difference in the masses of the \Dstarm and \Dzb candidates~($\Delta m$) must lie within 143 and $148\mevcc$.
Sideband regions of $m(\Dzb)$ in ranges $[1825, \,1840]\mevcc$ and $[1890,\, 1905]\mevcc$ and $\Delta m$ within $[150,\, 160]\mevcc$ are defined to study the combinatorial background.

The \taup candidates are formed from three tracks, where each track must have \mbox{$\pt>250\mevc$} and be consistent with the pion hypothesis. Additionally, the 3$\pi$ vertex must be separated from the PV associated with the signal decay by at least 10 times the uncertainty on the distance of separation. The radial distance between the $3\pi$ vertex and the beam center in the transverse plane is required to be within [0.2, 5.0]$\mm$ to avoid pion triplets coming from secondary interactions or a PV. The combinatorial background is suppressed by requiring that the \Dzb and \taup candidates are associated with the same PV. For events with $B^0$ mass greater than 5150~MeV/$c^2$, the significance of the IP of the \Dzb and $3\pi$ candidates with respect to the PV associated with the signal decay is required to be larger than 4.

\subsubsection{Particle identification requirements}
The charged pion and kaon tracks must be positively identified using the information provided by the particle identification~(PID) system. The requirements are the same as those used in the previous analysis~\cite{LHCb-PAPER-2017-017,LHCb-PAPER-2017-027}. Contamination due to decays of the type $\mbox{\decay{B}{\Dstarm D^+(X)}}$, with the subsequent decay $\mbox{\decay{D^+}{K^-\pi^+\pi^+(\pi^0)}}$, are suppressed by requiring that the kaon identification probability is less than 17\% for the $\pi^-$ candidate within the $3\pi$ system. 

\subsubsection{Anticombinatorial and isolation requirements}
\label{sec:anticomiso}

 The combinatorial background is further suppressed using a boosted decision tree~(BDT) classifier that is trained using simulated \sig decays as the signal proxy and the data sample with the same-sign charge combination  $\Dstarm \pim\pip\pim$ as the background proxy. The distributions of $\pt$ and $\eta$ of the \Dstarm and \taup candidates along with variables related to their vertex and IP are used to separate signal and background events. Two such variables are $\chi^2_{\text{IP}}$, which is defined as the difference in the vertex-fit $\chi^2$ of the PV reconstructed with and without the particle under consideration, and vertex $\chi^2$, which describes the quality of a vertex. The flight direction information of the $B^0$ and \taup is also utilized. The BDT classifier rejects about 75\% of the combinatorial background while preserving 77\% of the \sig decays.
 
 A closely related challenge is the rejection of partially reconstructed backgrounds with more than six charged tracks. The main source of these candidates is $\mbox{\decay{B}{\Dstarm\Dsp (X)}}$ processes where the \Ds meson decays into five stable charged 
 particles and $\mbox{\decay{B}{\Dstarm \Dz \Kp}}$ decays where the \Dz meson decays into four stable charged particles. A dedicated algorithm~\cite{LHCb-PAPER-2015-031} is used to evaluate the isolation of each signal-candidate track from other nonsignal tracks in an event. An isolation BDT classifier is formed from this information for each signal-candidate track, trained using simulated \sig decays as the signal proxy and simulated  $\mbox{\decay{B}{\Dstarm\Dz\Kp}}$ decays in which two extraneous charged kaons are present as the background proxy. This classifier removes 82\% of background decays with extra charged tracks while retaining 78\% of signal decays. The isolation requirements for suppressing background with extra neutral particles in the final state are discussed in \cref{sec:antiDs}.

 \subsection{Selection of signal mode \sig}
 
The signal-mode \sig candidates are identified with the use of a detached-vertex criterion and a targeted suppression of backgrounds where \Dsp decays mimic hadronic decays of \taup leptons.

\subsubsection{Vertex detachment criteria}
\label{sec:detachment}
In \sig decays, the 3\pion vertex is detached from the \Bz vertex. A good approximation for the \Bz vertex is the point of closest approach between the \Dstarm line of flight and the 3\pion line of flight. A BDT classifier is used to identify this detached topology. The inputs to this BDT classifier are the positions of the \Dzb, \Bz and 3\pion vertices and their related uncertainties, the 3\pion mass, and the momenta of the tracks forming the $\Bd$ candidate. The training of this BDT is performed using simulated \sig decays as signal and simulated prompt \mbox{$\bbbar \to \Dstarm 3\pi X$} production as background samples. 
The efficiency and rejection performance of this BDT classifier is slightly better than the beam-direction significance used in Refs.~\cite{LHCb-PAPER-2017-017,LHCb-PAPER-2017-027}; the rejection rate is 20\% higher for the same signal efficiency. 

\subsubsection{Suppression of \texorpdfstring{$\boldsymbol{\Ds}$}{Ds+} backgrounds}
\label{sec:antiDs}
The 3$\pi$ decay of the $\taup$ proceeds predominantly through an $a_1$ resonance with a $\rho^0\pi^+$ intermediate state. There is a  major background contribution due to \Dsp decays to 3$\pi(X)$, which primarily proceeds through $\eta$ and $\eta'$ resonances, with only a very small contribution from $R a_1$ structures, where $R$ designates an $\eta$, $\eta'$, $\omega$, $\phi$ or $\Kz$ meson~\cite{PDG2022}. In addition, for the latter three mesons, the phase space is such that the $a_1$ meson must be produced below its on-shell mass, providing further discrimination with respect to \tauon decays. 
The \pip\pim mass is kinematically limited to $<400\mevcc$ for the decays \mbox{$\eta\to\pip\pim\piz$} and \mbox{$\etapr\to\pip\pim\eta$}, which provides input to a discriminating variable formed by the minimum mass of the two \pip\pim combinations present in the pion triplet. The 3$\pi$ system from \Dsp decays is often accompanied with a large number of neutral particles from the same decay. Variables related to the energy from these neutral particles in cones around the $3\pi$ system direction are used in a BDT classifier to suppress these backgrounds. This classifier is trained using simulated \sig events as the signal training sample and simulated \Ds events decaying into three pions as the background training sample. The performance of this anti-\Ds BDT is better than that found in the previous analysis~\cite{LHCb-PAPER-2017-017,LHCb-PAPER-2017-027} with 40\% rejection of \Ds events for a 97\% signal efficiency. The anti-\Ds BDT output is used as a fit variable to estimate the yield of \sig decays.

\subsubsection{Other requirements}
The invariant mass of the 3$\pi$ system is required to be below 1600 \mevcc to suppress double-charm backgrounds.
An upper boundary for the invariant mass of the $D^{*}3\pi$ candidates is set at 5100~\mevcc, consistent with the presence of neutrinos in the final state of the signal decays. After all the selection requirements are applied, only 0.5\% of events have multiple candidates, from which one is chosen at random.

\subsection{Selection of normalization mode \texorpdfstring{$\boldsymbol{\norm}$}{B0->D*-3pi+/-}}

The selection of \norm candidates utilizes the unique characteristics of this mode: the fully reconstructed \mbox{$\Bd \to \Dstarm 3\pi$} decay and the $3\pi$ system coming directly from the \Bd meson. The \Bz candidates are selected in the mass range $[5150,5400]\mevcc$. The selection is kept similar to that of the \sig mode, to cancel the majority of systematic bias in the measurement of the ratio \kappadstar. To ensure this, the \Dzb decay vertex is required to lie further downstream than the $3\pi$ vertex, a similar detachment criterion to that for the $3\pi$ system from the $\Bz$ decay vertex in \sig decays. 
The anti-\Ds BDT classifier and $m(3\pi)$ selection criteria are not applied; however these do not bias \kappadstar, since their efficiency is around 97\% for the \sig decays.

\subsection{Simulation corrections and efficiencies}

The simulation samples used in the analysis are required to match the conditions in data as closely as possible. The \pt and $\eta$ distributions of the $B$ meson, \sig form factors, $3\pi$ vertex-position uncertainty and the $3\pi$ decay dynamics in simulation are calibrated to data. Control samples are used to validate simulation samples and apply corrections where necessary, as described in \cref{sec:doublecharm}. The selection efficiencies ($\varepsilon$) for the \sig and \norm modes are estimated from simulation after corrections are applied. The efficiencies used in Eq.~(\ref{eqn:kappa}) are $\varepsilon_{\rm sig} = (1.21 \pm 0.01) \times 10^{-4}$ and $\varepsilon_{\rm norm} = (3.54 \pm 0.04) \times 10^{-4}$. The uncertainties are due to the limited size of the simulation samples.

\section{STUDY OF DOUBLE-CHARM BACKGROUND}\label{sec:doublecharm}

The dominant background category after applying the selection criteria mentioned in Sec.~\ref{sec:selection} is double-charm decays $\mbox{\decay{B}{\Dstarm \D (X)}}$, where $D$ is a \Ds, $D^+$ or $D^0$ meson. Control samples in data are used to study these backgrounds and evaluate corrections that must be applied to simulated samples which are used to obtain the background probability density functions~(PDFs) used in the fit to determine the signal yield. The decays involving \Ds mesons are analyzed in two stages. First, the corrections to the branching fractions used to generate the simulation of different \Ds decays with three pions in the final state are estimated. These corrections are applied to the simulation samples. Second, the composition of several $\mbox{\decay{B}{\Dstarm\Ds (X)}} $ decays is determined, serving as constraints on the fractions of these components in the signal-extraction fit.

\subsection{\texorpdfstring{$\boldsymbol{\Ds \to 3\pi X}$}{ Ds+->3piX} decay model}
\label{dscecay}
The decays of \Ds mesons and \taup leptons to final states involving three pions are distinct, as discussed in \cref{sec:antiDs}. However, the \Ds meson can decay to the $\rho^0$ meson via an $\eta'$ resonance, which decays to the $\rho^0\gamma$ final state. Consequently, the $\rho^0$ contribution from the \Ds decay could be mistakenly attributed to that from a \taup decay. Therefore, it is crucial that contributions from various resonances, especially $\eta'$, are correctly normalized in simulation to reflect the data as closely as possible. It is also essential to constrain the relative contributions of certain decay modes, whose branching fractions are not precisely measured.

The fractions of different \mbox{$\Ds \to 3\pi X$} decays are determined from a data sample enriched in these decays. This sample is selected with the same criteria as for \sig decays but with a reverse requirement on the anti-\Ds BDT output. A simultaneous binned maximum-likelihood fit is performed to the distribution of four variables: 
$\text{min}[m(\pip \pim)]$, $\text{max}[m(\pip \pim)]$, $m(\pip \pip)$, which represents the reconstructed masses of all possible two-pion combinations of the candidate, and $m(3\pi)$. This last variable allows the exclusive \mbox{$\Dsp\to 3\pi$} decay to be distinguished from candidates where energy is carried by additional
neutral particles, \eg,~\mbox{$\Dsp~\to~\taup~(\to~3\pi~\neutb)$~\neut} or \mbox{$\Dsp~\to~3\pi~\PX$}, where \PX escapes detection.

The different \Ds decay components are broadly divided into four categories:
\begin{itemize}
 \item $\Dsp \to \eta \pip (\piz)$ decays where charged pions from the $\eta$ meson are selected,
 \item $\Dsp \to \eta' \pip (\piz)$ decays where charged pions from the $\eta'$ meson are selected,
 \item $\Dsp \to \omega \pip (\piz)$ or $\Dsp \to \phi \pip (\piz)$  decays where charged pions from the $\omega$ or $\phi$ meson are selected,
 \item \Dsp decays where the pions originate either directly from the \Dsp decay or from the $a_1$ resonance $\eta 3\pi$, $\eta a_1$, $\eta '3\pi$, $\eta' a_1$, $\omega 3\pi$, $\omega a_1$, $\phi 3\pi$, $\phi a_1$, $K^0 3\pi$, $K^0 a_1$, $\taup \neut$ and nonresonant $3\pi$.
\end{itemize}
\begin{figure}[t]
  \centering
   \includegraphics[width=.49\linewidth]{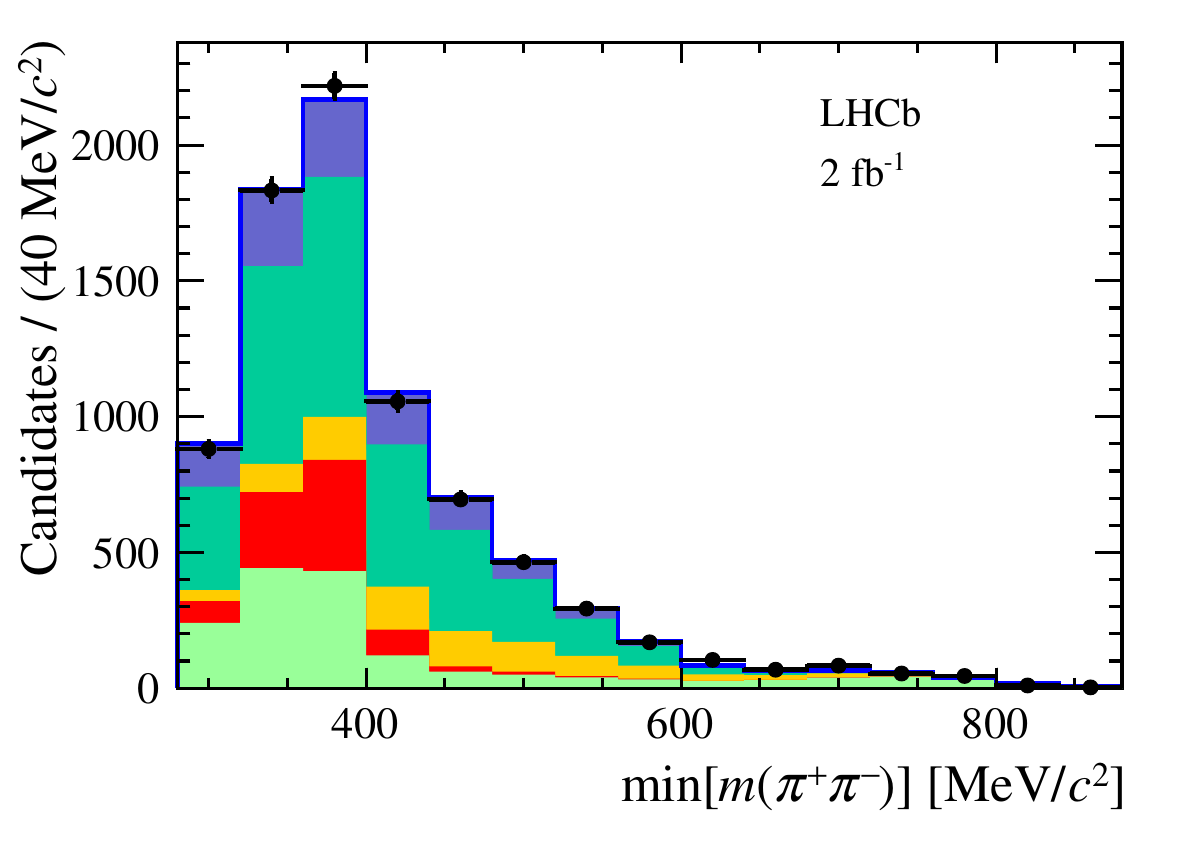}
  \includegraphics[width=.49\linewidth]{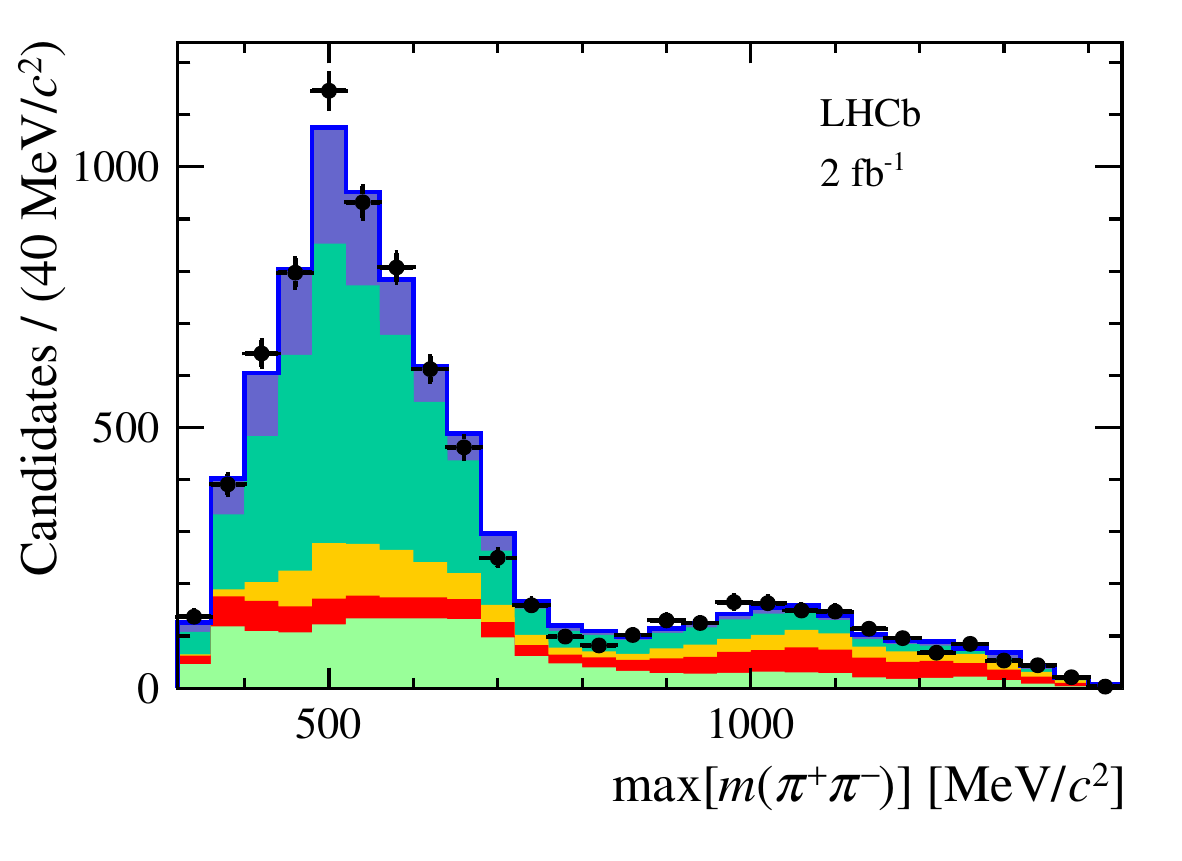}
  \includegraphics[width=.49\linewidth]{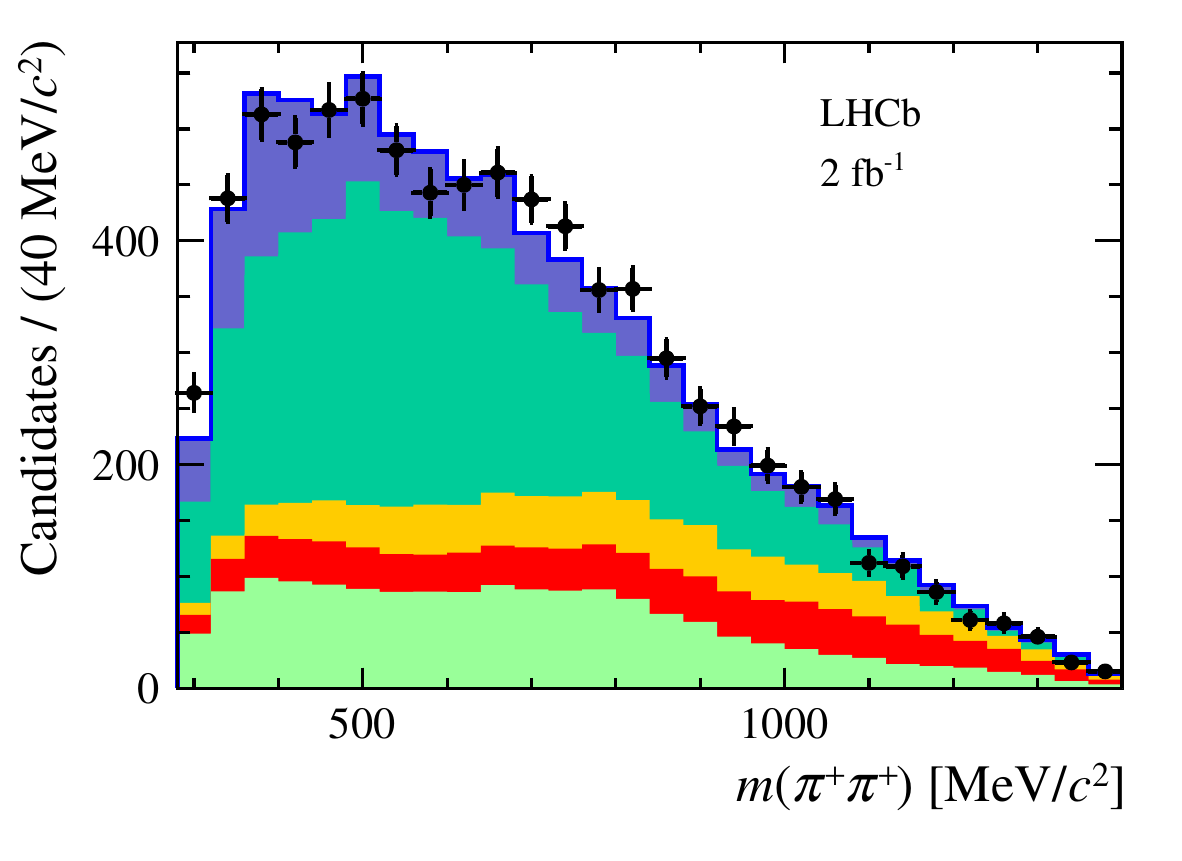} 
  \includegraphics[width=.49\linewidth]{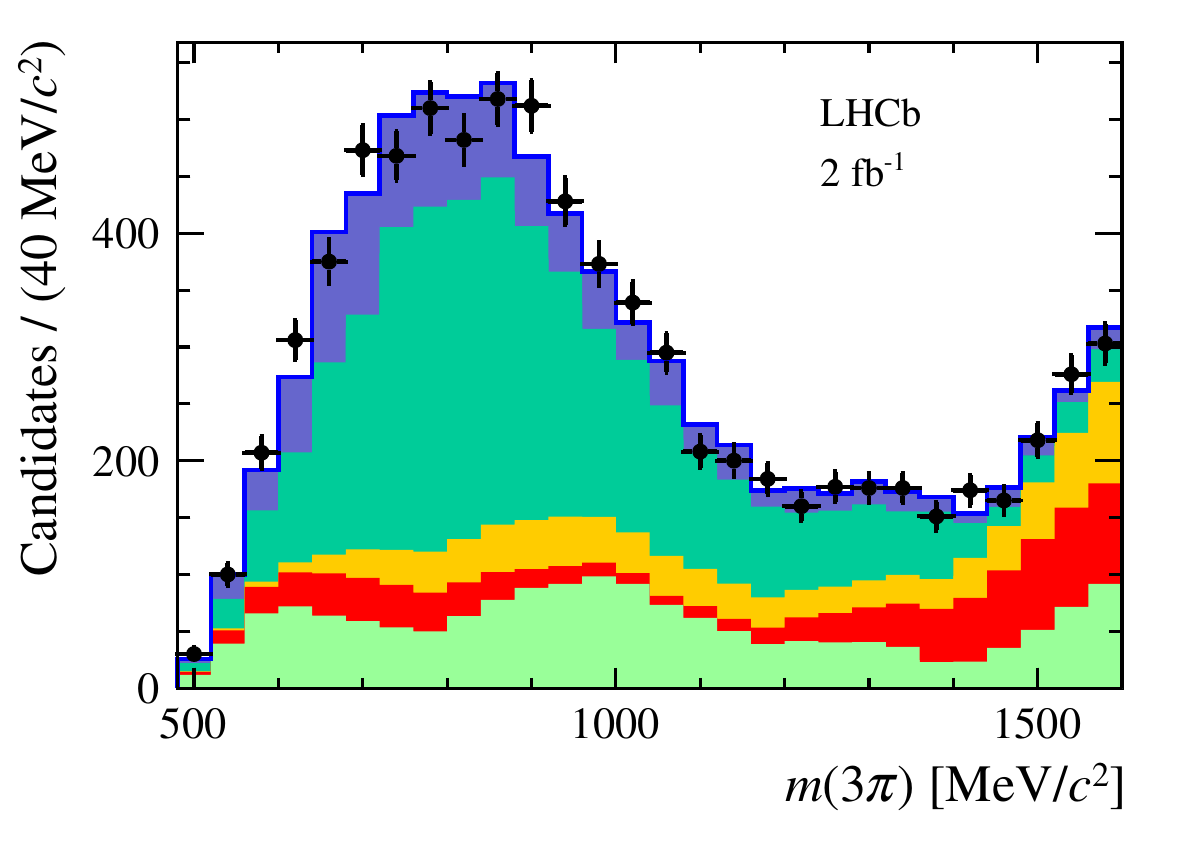}
\begin{minipage}[c]{1.0\textwidth}
    \hspace{45pt}
    \includegraphics[width=0.9\linewidth]{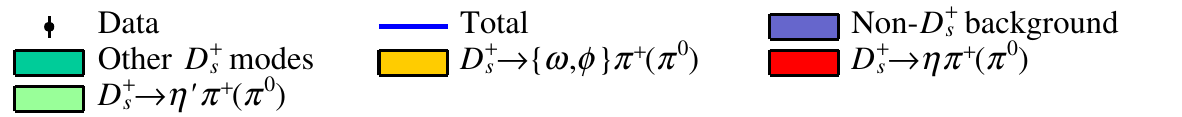}
\end{minipage}
  \caption{Projections of the $\Dsp \to 3\pi \PX$ components for the variables: min[$m(\pip \pim)$], max[$m(\pip \pim)$], $m(\pip \pip)$ and $m(3\pi)$
  in the fit to the control data samples. 
  }
  \label{fig:dspfit}
\end{figure}
The template PDFs for the various \Dsp decay components are defined using inclusive \mbox{$\decay{B}{\Dstarm\Ds X}$} simulation samples, and the non-\Ds decays are modeled using the inclusive \mbox{$B \to \Dstarm 3\pi X$} simulation sample.  The fraction of each of the four \Dsp decay components, relative fractions of $\eta 3\pi$, $\eta' 3\pi$ and $\omega 3\pi$ final states and the total number of \Dsp and non-\Dsp decays are free parameters in the fit. Compared to the previous analysis~\cite{LHCb-PAPER-2017-017,LHCb-PAPER-2017-027}, the simulated samples are improved with the addition of \mbox{$\Ds \to R a_1$} modes with $R \in \{ \eta, \eta', \omega, \phi, K^0\}$, which enable a more detailed description of the $\Dsp\to 3\pi X$ decay. This introduces more fit parameters and hence additional constraints are applied to ensure the stability of the fit. The fit assumes no interference effects, as a full amplitude analysis is beyond the scope of this study. The fit results are given in \cref{fig:dspfit}, which shows the fractions of the four inclusive categories mentioned above. The relative fraction of the component with the $\piz$ meson with respect to the total fraction of each of the first three categories are Gaussian constrained around the expected value of $2/3$ with a standard deviation of $0.2$. For each resonance $R$, the sum of \mbox{$\Ds \to Ra_1$} and \mbox{$\Ds \to R3\pi$} is measured. The relative fraction \mbox{$\Ds \to Ra_1$}/(\mbox{$\Ds \to Ra_1 + \Ds \to R3\pi$}) is fixed at 5.5\%. This is chosen since it provides the best quality for the signal fit. The fractions of \Ds decays to $ \phi a_1+ \phi 3\pi$, $K^0a_1+K^03\pi$ and $\taup \neut$ final states are fixed according to their known branching fractions~\cite{PDG2022}.

\begin{table}[t]
\centering
\caption{Relative fractions of various \mbox{$\Dsp \to 3\pi \PX$} decays obtained from the fit to the $\Dsp$ control sample (see~\cref{fig:dspfit}) before applying any bias correction to the mean value of its uncertainty.}
\begin{tabular}{lc}
\hline
 Template & $\Dsp \to 3\pi \PX$ fraction \\ \hline
 $\eta \rhop$ & $0.112 \pm 0.020$  \\
$\eta \pip$ & $0.021 \pm 0.008$  \\
$\etapr \rhop$ & $0.185 \pm 0.016$  \\
$\etapr \pip$ & $0.051 \pm 0.014$  \\
$\omega \rhop \text{ or } \phi \rhop$ & $0.090 \pm 0.029$  \\
$\omega \pip \text{ or } \phi \pip$ & $0.041 \pm 0.008$  \\
$\eta 3\pi$ & $0.081 \pm 0.024$  \\
$\etapr 3\pi$ & $0.000 \pm 0.005$  \\
$\omega 3\pi$ & $0.032 \pm 0.041$  \\
$\phi 3\pi$ & $0.029 \pm 0.008 $  \\
$\Kz 3\pi$ & $0.011 \pm 0.003$ \\
$\taup \neut$ & $0.012 \pm 0.003$  \\
Nonresonant $3\pi$ & $0.326 \pm 0.046$  \\
$a_1 \eta$ & $0.005 \pm 0.001$  \\
$a_1 \etapr$ & $0.000 \pm 0.001$  \\
$a_1 \omega$ & $0.002\pm 0.002$ \\
$a_1 \phi$ & $0.002 \pm 0.001$  \\
$a_1 \Kz$ & $0.001 \pm 0.001$  \\
\hline
\end{tabular}
\label{tab:DspParams}
\end{table}

The $\chi^2$ per degree of freedom is evaluated to be 1.5, indicating a reasonable fit quality. Possible effects from mismodeling of the components in the fit are considered as sources of systematic uncertainties, which are described  in \cref{Sec:syst}.
The determined fractions of the different modes are given in \cref{tab:DspParams}. They are used to correct the corresponding modes in the simulation sample. The uncertainties obtained from the fit could be underestimated due to the statistical correlations arising from simultaneously fitting to four one-dimensional distributions.  The effect of the correlations is investigated using a set of pseudodata samples built via bootstrapping from simulation, in which events are assigned to each component category randomly. The statistical uncertainty on the fit parameters is found to be underestimated by no more than 40\%. The uncertainties of the fit parameters are then corrected according to this estimation. The correction factors are applied to the simulation samples that are used to produce the template PDFs for \mbox{$\decay{B}{\Dstarm\Ds X}$} backgrounds.

\subsection{\texorpdfstring{$\boldsymbol{\decay{B}{\Dstarm\Ds (X)}}$}{B->D*-Ds+(X)} control sample}\label{sec:dstdsxfit}
The relative abundance of different \mbox{$\decay{B}{\Dstarm\Ds (X)}$} decays provides important constraints in the signal fit. These relative fractions are determined from a fit to the data control sample enriched in \mbox{\decay{\PB}{\Dstarm\Ds(\to 3\pi)(X)}} decays. The selection of the control sample differs from the default selection by requiring the $3\pi$ mass to be within 20\mevcc of the known \Ds mass
and omitting the anti-$\Dsp$ BDT cut, as well as the $\Bd$ and $\taup$ mass constraints. 
This sample comprises modes that can be grouped into the exclusive \mbox{$\Bd \to \Dstarm D_s^{(*,**)+}$} decays, and the inclusive \mbox{$\B \to {\Db}^{**} \Dsp (X)$} and \mbox{$\Bs \to \Dstarm \Dsp (X)$} decays.\footnote{Throughout the paper, ${\Db}^{**}$ and $D_s^{**+}$ are used to refer to any higher-mass excited charm or charm-strange mesons that decay into the ground-state \Dstarm and \Dsp meson.}

The $q^2 \equiv (p_{\Bd} - p_{\Dstarm})^2$ distribution for exclusive modes peaks at the mass of the relevant $D_s^{(*,**)+}$ states, where $p_H$ is the momentum of particle $H$. 
The inclusive \mbox{$B \to \Dstarm \Ds \PX$} modes often have at least one extra particle, possibly from $\Db^{**}\Ds (X)$ final states.
These additional particles carry momentum that contributes to the $q^2$, shifting this distribution to higher values. 

An extended binned maximum-likelihood fit is performed to the distribution of the difference of the $\Dstarm 3\pi$ mass and the sum of the reconstructed \Dzb and $3\pi$ masses, \ie,\ \mbox{$\Delta M _{D^*D_s}\equiv m(\Dstarm 3\pi)-m(\Dzb)-m(3\pi)$}.
The total PDF used in the fit is
\begin{equation}
 \mathcal{P} = f_{\text{comb}} \mathcal{P}_{\text{comb}} + \frac{(1-f_{\text{comb}})}{k} \sum_i f_i \mathcal{P}_i,
\end{equation}
where $f_{\text{comb}}$ is the fraction of the combinatorial background and $\mathcal{P}_{\text{comb}}$ is the corresponding PDF that is modeled using the same-sign $\Dstarp 3\pi$ data sample; $f_i$ are floating fractions of the different \mbox{$B \to \Dstarm \Dsp (X)$} components with \mbox{$i\in\{ \Ds\,, \Dssz\,, \Dsprime\,, \Db^{**}\Dsp (X)\,, B_{s}^0~\to~\Dstarm \Ds (X)\,, \Dss \}$}, relative to the most abundant \mbox{$\Bd \to \Dstarm \Dss$} decays; and
$k = 1 + \sum_i f_i$ by definition. Here, $\Dssz$ and $\Dsprime$ denote the $D_{s0}^*(2317)^+$ and $D_{s1}(2460)^+$ states, respectively.
The template shape of each component is taken from simulation.

\begin{figure}[tbp]
  \centering
    \includegraphics[width=.49\linewidth]{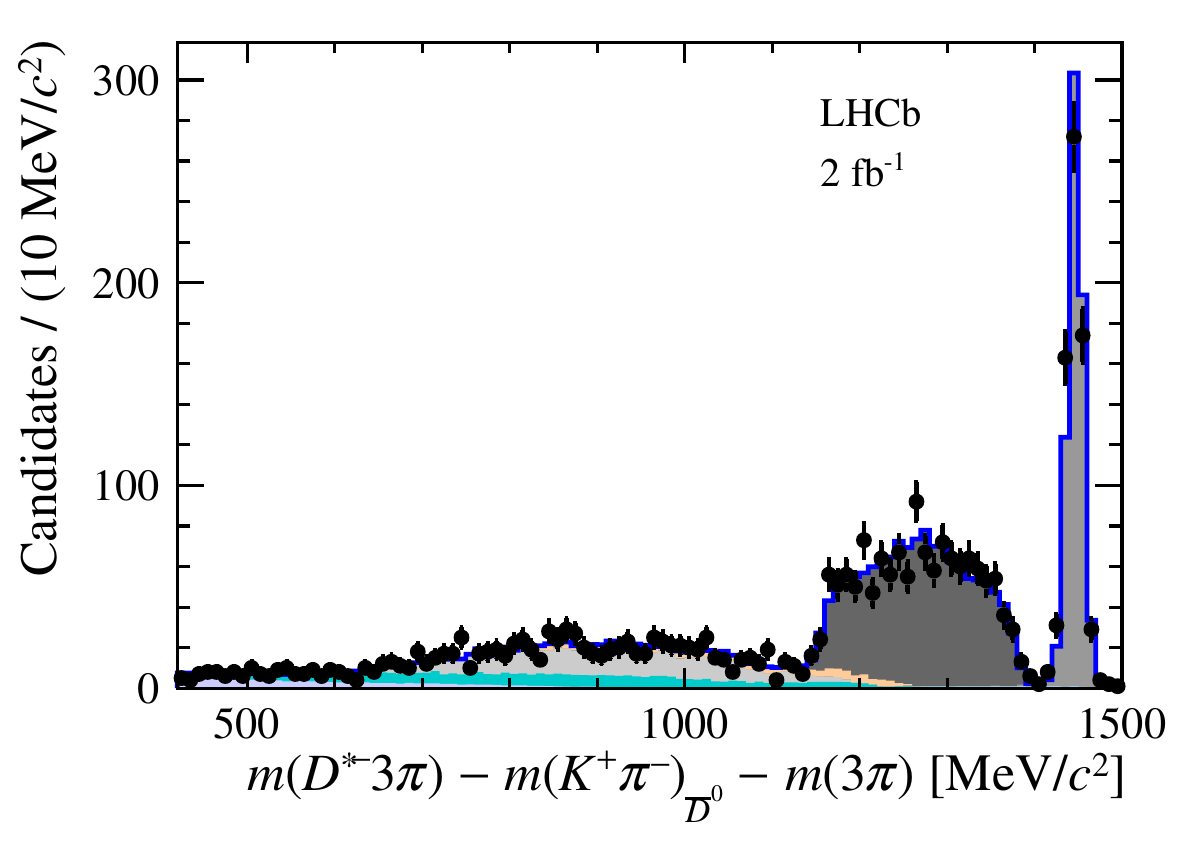}
\includegraphics[width=.49\linewidth]{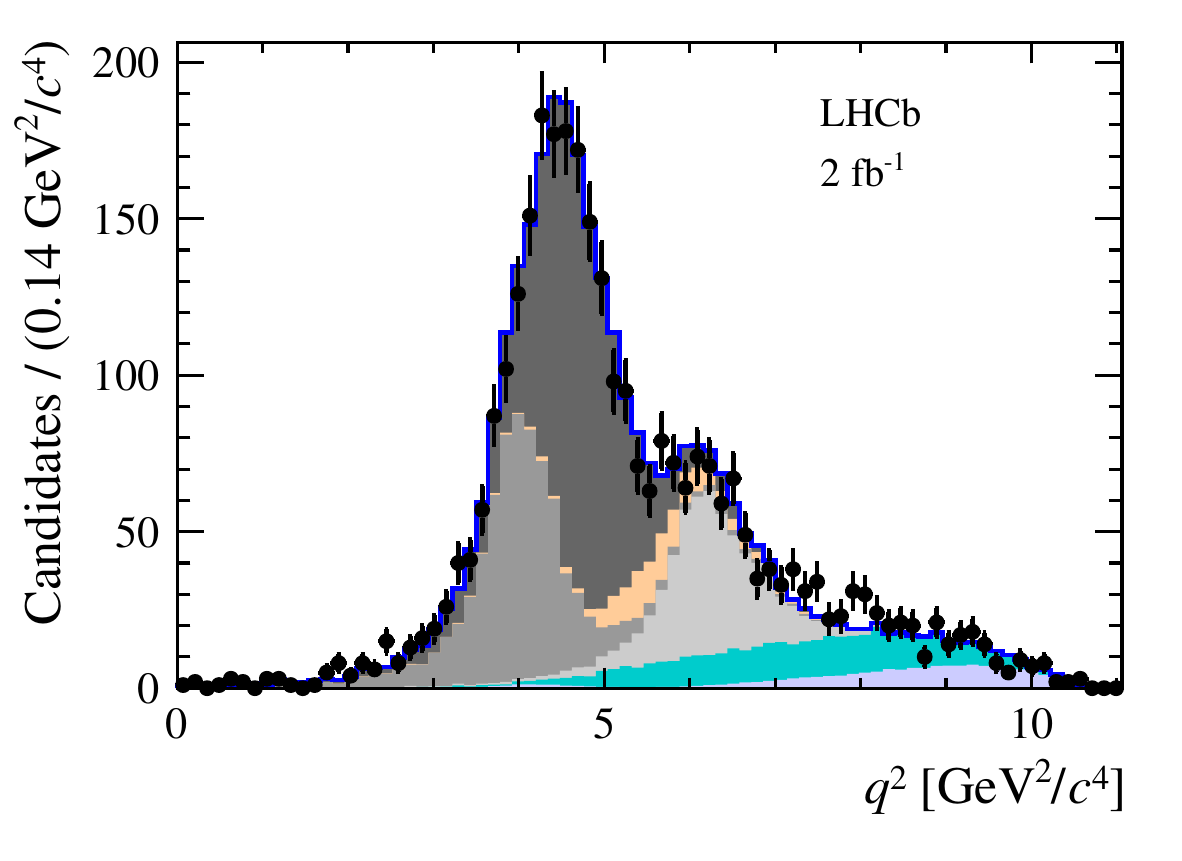}
  \includegraphics[width=.49\linewidth]{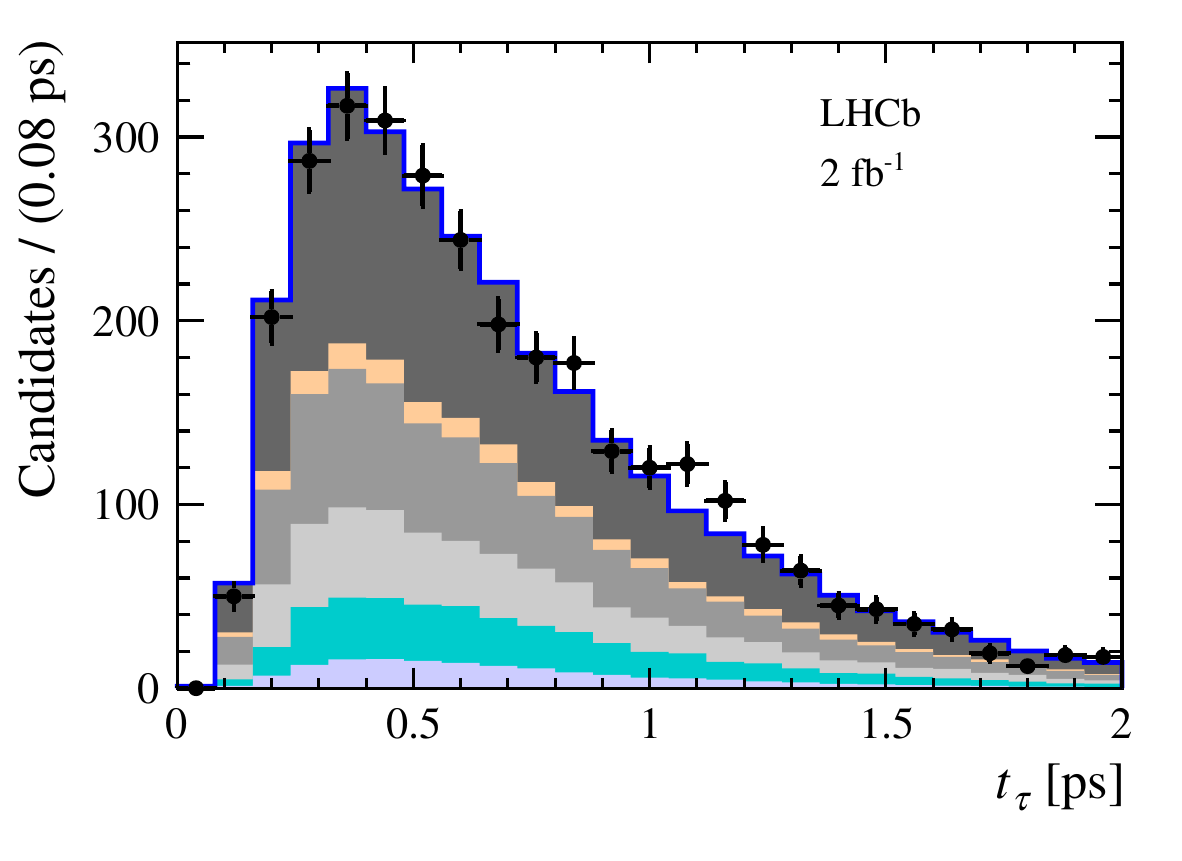}
  \includegraphics[width=.49\linewidth]{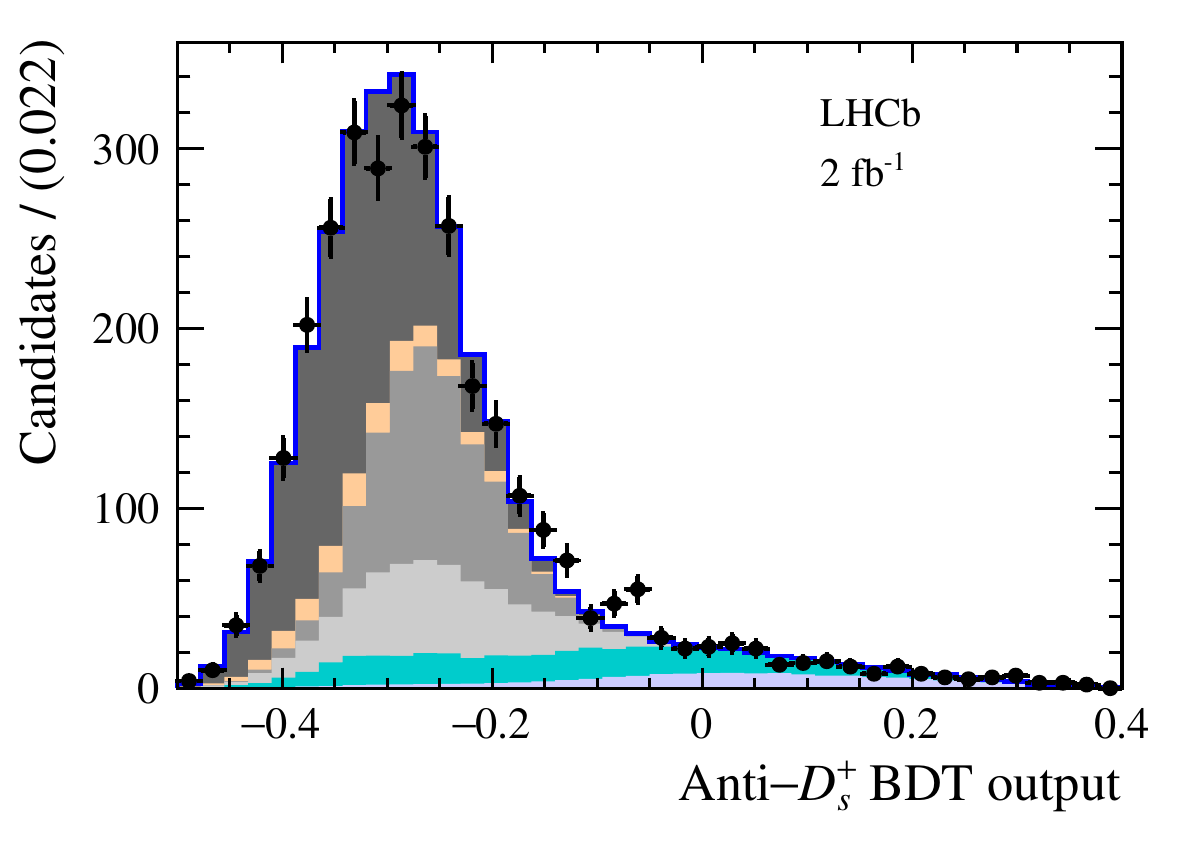}
  \begin{minipage}[c]{1.0\textwidth}
    \hspace{45pt}
    \includegraphics[width=0.9\linewidth]{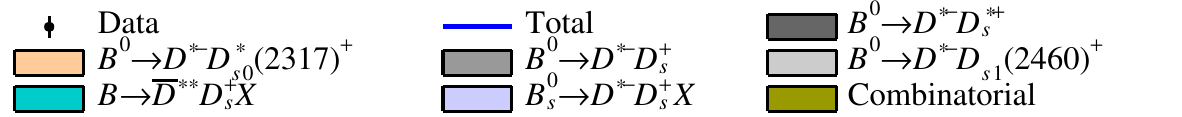}
\end{minipage}
  \caption{Distributions of $\Delta M _{D^*D_s}$, $q^2$, $t_{\tau}$ and anti-\Dsp BDT output for the \mbox{$\PB \to \Dstarm \Dsp (X)$} components. The results of the fit are overlaid.}
  \label{fig:BtoDstDsX}
\end{figure}

The distributions of $\Delta M _{D^*D_s}$, $q^2$, decay time of the \taup candidate ($t_{\tau}$) and anti-\Dsp BDT output are shown in \cref{fig:BtoDstDsX}. The fit quality is good with a $\chi^2$ per degree of freedom equaling 1.11. 
The fractions of different decays are given in \cref{tab:dstdsx}, which are used as Gaussian constraints in the signal-extraction fit for the corresponding components after accounting for the efficiency differences between the control and signal samples. 

\begin{table}[tbp]
    \centering
    \caption{Decay fractions for \mbox{$\decay{B}{\Dstarm\Ds (X)}$} decays obtained from data control samples. The fractions are normalized relative to that of the \mbox{$\Bd \to \Dstarm \Dss$} decay. $\epsilon_{\textnormal{control}}$ is the efficiency in the control sample.}
    \label{tab:dstdsx}
    \begin{tabular}{lccc}
    \hline 
    Parameter & Fit result & $(\frac{\epsilon_{\textnormal{sig}}}{\epsilon_{\textnormal{control}}})$ & Corrected fraction\\[0.5ex] \hline 
    $ f_{\Dsp} $ & $0.55 \pm 0.03$ & 0.992 & $0.55 \pm 0.03$ \\
    $f_{D_{s0}^{*+}}$ & $0.10 \pm 0.04$ & 1.077 &$0.11 \pm 0.04$\\
    $f_{D_{s1}^{+}}$ & $0.37 \pm 0.07$ & 1.051 & $0.39 \pm 0.07$ \\
    $f_{{\Db}^{**} \Dsp (X)}$ & $0.28 \pm 0.10$ & 1.208 & $0.34 \pm 0.12$\\
    $ f_{\Bs \rightarrow \Dstarm \Dsp (X)}$ & $0.12 \pm 0.04$& 0.904 & $0.11 \pm 0.04$ \\[0.5ex]
    \hline
\end{tabular}
\end{table}

\subsection{\texorpdfstring{$\boldsymbol{\decay{B}{\Dstarm(D^0,D^+) (X)}}$}{B->D*-(D0,D+) (X)} control samples}
The \mbox{$B \to \Dstar^-\D^+(X)$} and \mbox{$B \to \Dstar^-\D^0(X)$} decays are the subleading double-charm backgrounds in the signal sample, where the $D^0$ mesons decay to three charged pions plus extra particles and the $D^+$ mesons decay to the $\pi^+K^-\pi^+$ final state with the kaon misidentified as a pion. Data control samples are used to check the agreement with simulation. 
A control sample  representing the \mbox{$\B \to \Dstarm \Dz \PX$} decays is selected using the decay mode \mbox{$\Dz \to \Km 3\pi$}. 
The isolation algorithm described in \cref{sec:anticomiso} searches for extra kaons around the $3\pi$ vertex and thus can be used to select such candidates.
A \mbox{$\B \to \Dstarm \Dp \PX$} control sample is obtained by reversing the PID requirements on the negatively charged pion to form \mbox{$\Dp\to \pip \Km \pip$} decays. In order to retain high statistics, the BDT and \Bd and \taup mass requirements are omitted. Other selection criteria remain the same as those for \sig decays.

\begin{figure}
    \centering
    \includegraphics[width=0.49\linewidth]{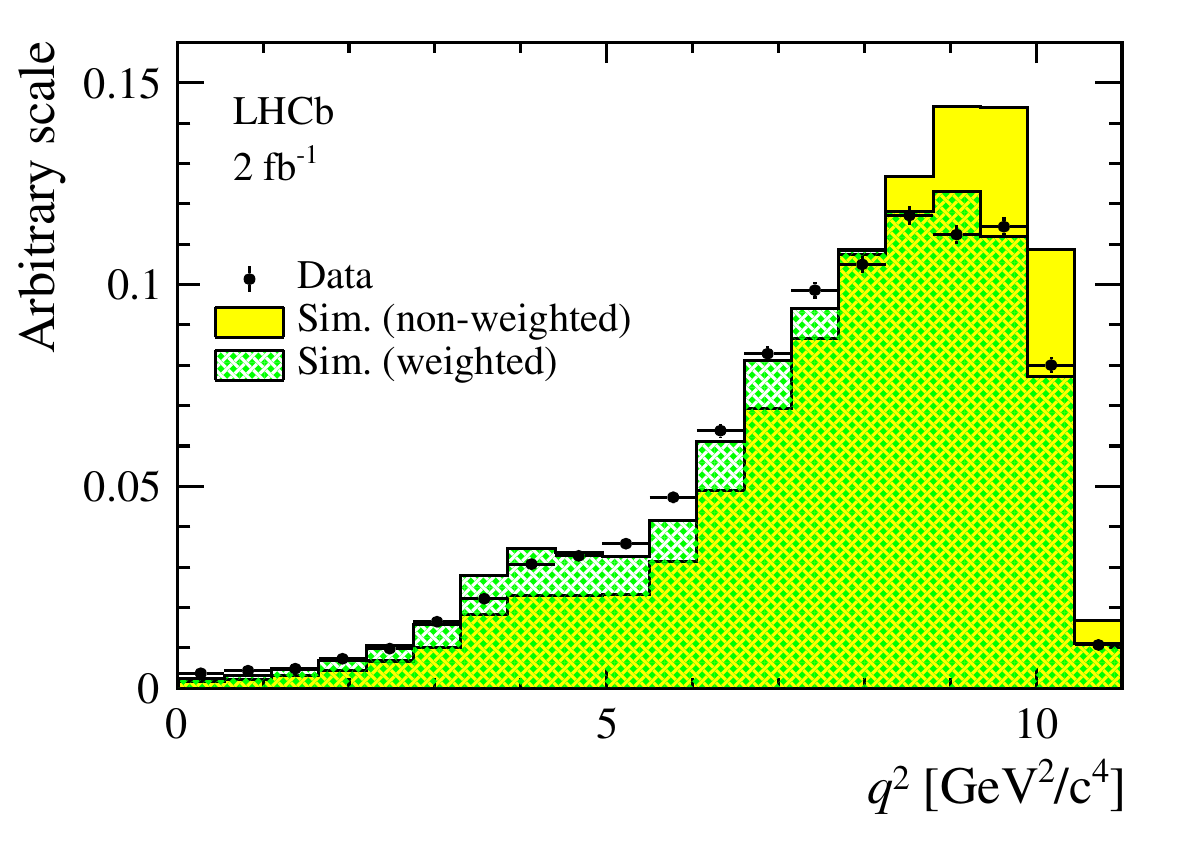}
    \caption{Simulated $q^2$ distributions for  \mbox{$B \to D^{*-}D^0(X)$} before and after weighting 
    based on the data control samples.} 
	  \label{fig:q2_correction}
\end{figure}

These control samples are used to check the agreement with simulation for the signal fit variables $q^2$, $t_{\tau}$ and anti-\Ds BDT output. The $q^2$ distribution shows disagreement between data and simulation in both $D^0$ and $D^+$ modes due to the imperfect modeling of inclusive $3\pi$ decays. Therefore, the simulation is corrected to match the data distributions. \Cref{fig:q2_correction} shows the $q^2$ distributions before and after the corrections in the \mbox{$\decay{B}{\Dstarm D^0 (X)}$} sample. The agreement is good in the case of the other two fit variables, $t_{\tau}$ and anti-\Ds BDT output, and no further correction is necessary.
\section{DETERMINATION OF THE SIGNAL AND NORMALIZATION YIELDS}
\label{sec:yield}
\subsection{\texorpdfstring{$\boldsymbol{\sig}$}{B0->D*-tau nv} yield}

The \sig yield is determined from an extended binned maximum-likelihood fit to the distributions of $q^2$, $t_\tau$ and anti-$\Dsp$ BDT output. The binning scheme comprises eight bins in $q^2$ and $t_{\tau}$, and six bins in the BDT output. The chosen binning scheme has maximum sensitivity to the \sig yield while still sufficiently populating the bins. The ranges of $q^2$, $t_{\tau}$ and BDT distributions are $[0,11]$~GeV$^2$/$c^4$, $[0,2]$~ps and $[-0.2,0.5]$, respectively. The fit model is built as a three-dimensional template.  A summary of the components in the fit model is given in \cref{tab:sigfitnormpars}. The fit model assumes that any possible new physics effects on the \sig decays are the same as that in \feeddown decays. The templates for the combinatorial components are derived from data in which same-sign combinations of the $D^*$ and $\tau$ candidates are selected, whereas the remainder are obtained from simulation. 

\begin{table}[t]
	\caption{List of components in the signal yield extraction fit and their normalization.}
	\label{tab:sigfitnormpars}
	\centering
        \begin{tabular}{l|l}
        Component                                      & Normalization \\
        \hline
        \signalmode (\tauthreepi)                      & $ N_{\text{sig}} \times f_{\tauthreepi} $ \\
        \signalmode (\tauthreepipiz)                   & $ N_{\text{sig}} \times (1 - f_{\tauthreepi}) $ \\
        \feeddown                                      & $ N_{\text{sig}} \times f_{\Db^{**} \taup \nu} $ \\
        \hline
        \decay{\PB}{\Dstarm \Dz (X)} same vertex (SV)      & $ N_{\Dz}^{\text{same}} $ \\
        \decay{\PB}{\Dstarm \Dz (X)} different vertices (DV)  & $ N_{\Dz}^{\text{same}} \times f_\Dz^{v_1v_2} $ \\
        \hline
        \decay{\PB}{\Dstarm \Dp (X)}                   & $ N_\Ds \times f_\Dp $ \\
        \hline
        
        \decay{\Bz}{\Dstarm \Ds}                       & $ N_\Ds \times f_{\Ds} / k $ \\ \hline
        \decay{\Bz}{\Dstarm \Dss}                      & $ N_\Ds \times 1 / k $ \\
        \decay{\Bz}{\Dstarm \Dssz}                     & $ N_\Ds \times f_{\Dssz} / k $ \\
        \decay{\Bz}{\Dstarm \Dsprime}                  & $ N_\Ds \times f_{\Dsprime} / k $ \\ \hline
        \decay{\PB}{\Db^{**} \Ds (X)}               & $ N_\Ds \times f_{\Db^{**} \Dsp (X)} / k $ \\
        \decay{\Bs}{\Dstarm \Ds (X)}                   & $ N_\Ds \times f_{\Bs \to \Dstarm \Ds (X)} / k $ \\
        \hline
        \decay{\PB}{\Dstarm 3\pi \PX}                & $ N_{\decay{\PB}{\Dstarm 3\pi \PX}} $ \\
        \hline
        Combinatorial \PB                               & $ N_{\PB_1\PB_2} $ \\
        Combinatorial \Dzb                            & $ N_{\text{fake \Dzb}} $\\
        Combinatorial \Dstarm                            & $ N_{\text{fake \Dstarm}} $\\ \hline
\end{tabular}
\end{table}
The fit parameters are itemized below.

\begin{itemize}
\item $N_{\text{sig}}$: the number of \signalmode events, which is used as input to \KDstar.
  
\item $f_{\tauthreepi}$: the fraction of \tauthreepi decays relative to the sum of \tauthreepi and \tauthreepipiz decays. This is estimated and fixed as per the branching fractions and efficiencies of these modes.
          
\item $f_{\Db^{**} \tau \nu}$: the amount of \feeddown decays relative to \signalmode decays. This is fixed to the expected value from simulation after correcting for the overestimated branching fractions used to produce them. The correction is done by comparing the theoretical expectations for each $\Db^{**}$ state from the $R(D^{**})$ predictions from Ref.~\cite{bernlochner2021semitauonic} and branching fractions of \mbox{$B \to \bar{D}^{**}\mu^+\nu_{\mu}$}~\cite{PDG2022}. This fraction is determined to be 3.5\%, which is significantly lower than that used in Refs.~\cite{LHCb-PAPER-2017-017,LHCb-PAPER-2017-027}.

\item $N_{\Dz}^{\text{same}}$: the number of \mbox{\decay{\PB}{\Dstarm \Dz (X)}} candidates where all pions in the $3\pi$ system originate from the \Dz vertex. The yield is estimated from simulation and corrected for data-simulation differences by the ratio of \mbox{$D^0 \to K^-3\pi$} decays in both samples. This value is fixed in the fit.

\item ${f_{D^{0}}^{v_1v_2}}$: the ratio of the number of \mbox{$B\to D^{*-}D^0 (X)$} decays where
at least one of the pions comes from the $D^0$ vertex and the other pion(s) from a different vertex relative to ${N_{D^{0}}^{\text{same}}}$. 

\item ${f_{D^+}}$: the ratio of the number of \mbox{$B\to \Dstarm D^+ (X)$} decays to the number of \mbox{$\decay{B}{\Dstarm\Ds (X)}$} decays.

\item $N_{\Dsp}$: the yield of \mbox{$\decay{B}{\Dstarm\Dsp (X)}$} decays, which have six categories, as described in \cref{sec:dstdsxfit}. The fraction parameters are Gaussian constrained to the values obtained from the data control sample given in Table~\ref{tab:dstdsx} after correcting for efficiency effects. 

\item ${N_{B\to \Dstarm 3\pi X}}$: the yield of \mbox{$B\to D^{*-}3\pi X$} events
where the three pions come from the $B$ vertex. This value is constrained by
using the observed ratio between \mbox{$B^0\to D^{*-}3\pi$} exclusive and \mbox{$B\to D^{*-}3\pi X$}
inclusive decays, corrected for data-simulation differences.
\item ${N_{B_{1}B_{2}}}$: the yield of combinatorial background events
where the $\Dstarm$ meson and the $3\pion$ system come from different $B$ decays. It is fixed to the value obtained in the same-sign data sample with $\Dstarp 3\pi$ candidates satisfying the criteria of higher mass and nonisolation due to originating from two different $b$ hadrons.

\item ${N_{\text{fake } \Dzb}}$ and ${N_{{\text{fake }} \Dstarm}}$: the combinatorial background yields
with a fake \Dzb and \Dstarm, respectively. These are fixed to the values obtained from a fit to $m(\Km \pip)$ and $m(\Dstarm) - m(\Km \pip)$.
\end{itemize}

\begin{figure}[t]
  \centering
\begin{minipage}[t]{.495\textwidth}
    \vspace{0pt}
    \centering
    \includegraphics[width=1.0\linewidth]{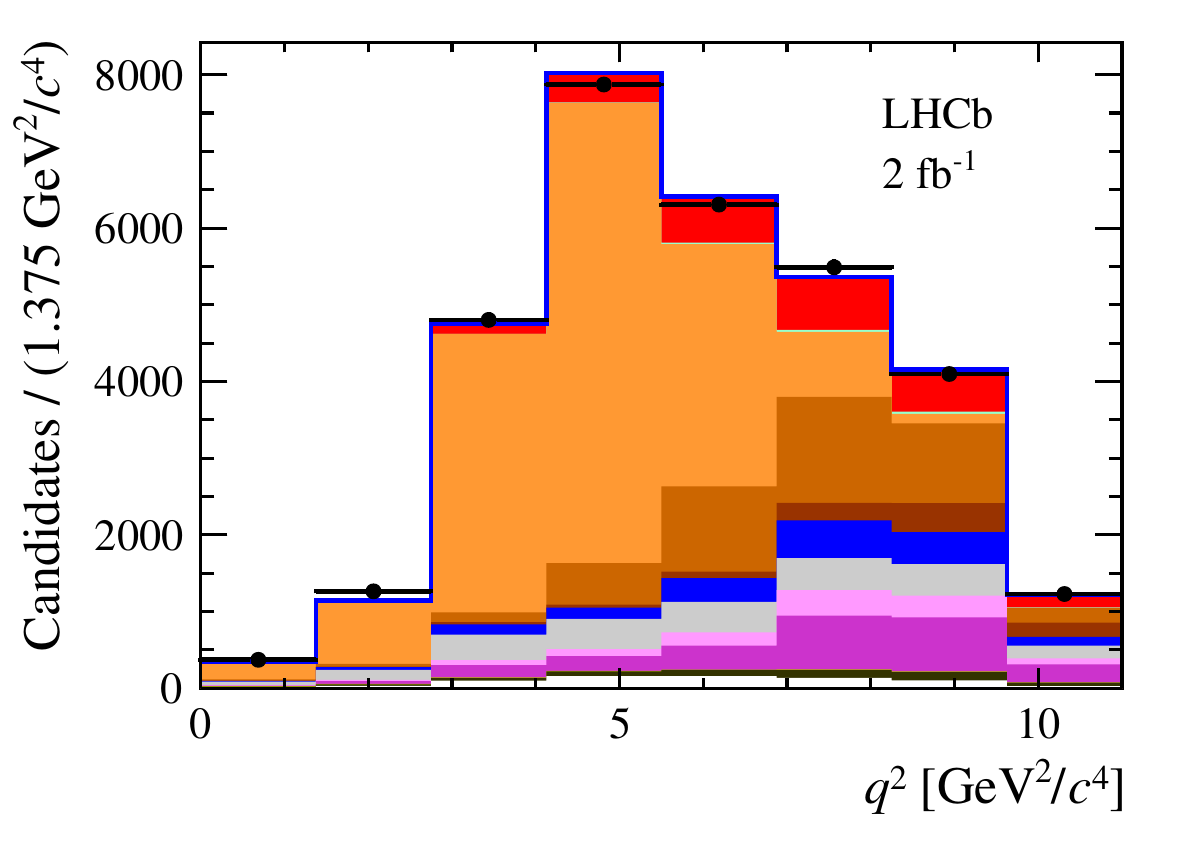}
\end{minipage}
\begin{minipage}[t]{.495\textwidth}
    \vspace{0pt}
    \centering
    \includegraphics[width=1.0\linewidth]{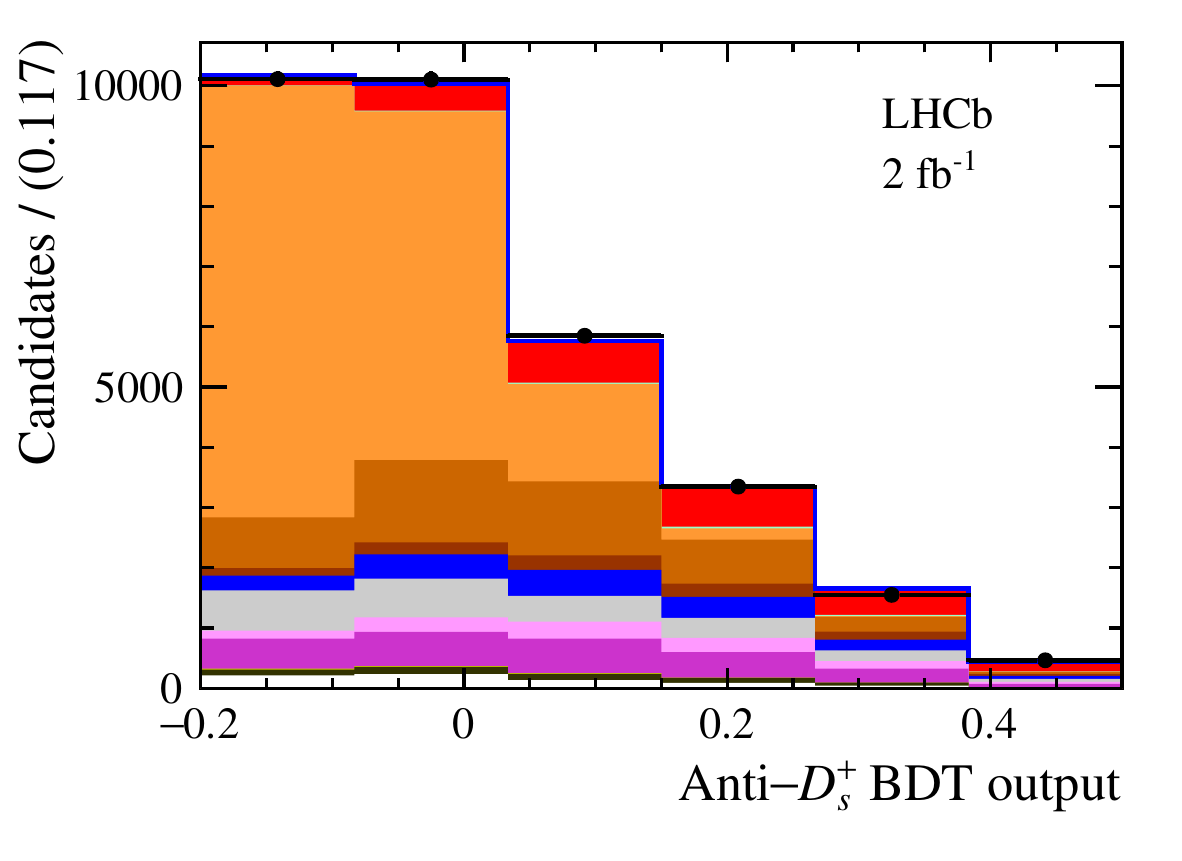}
\end{minipage}
\begin{minipage}[t]{.495\textwidth}
    \vspace{0pt}
    \centering
    \includegraphics[width=1.0\linewidth]{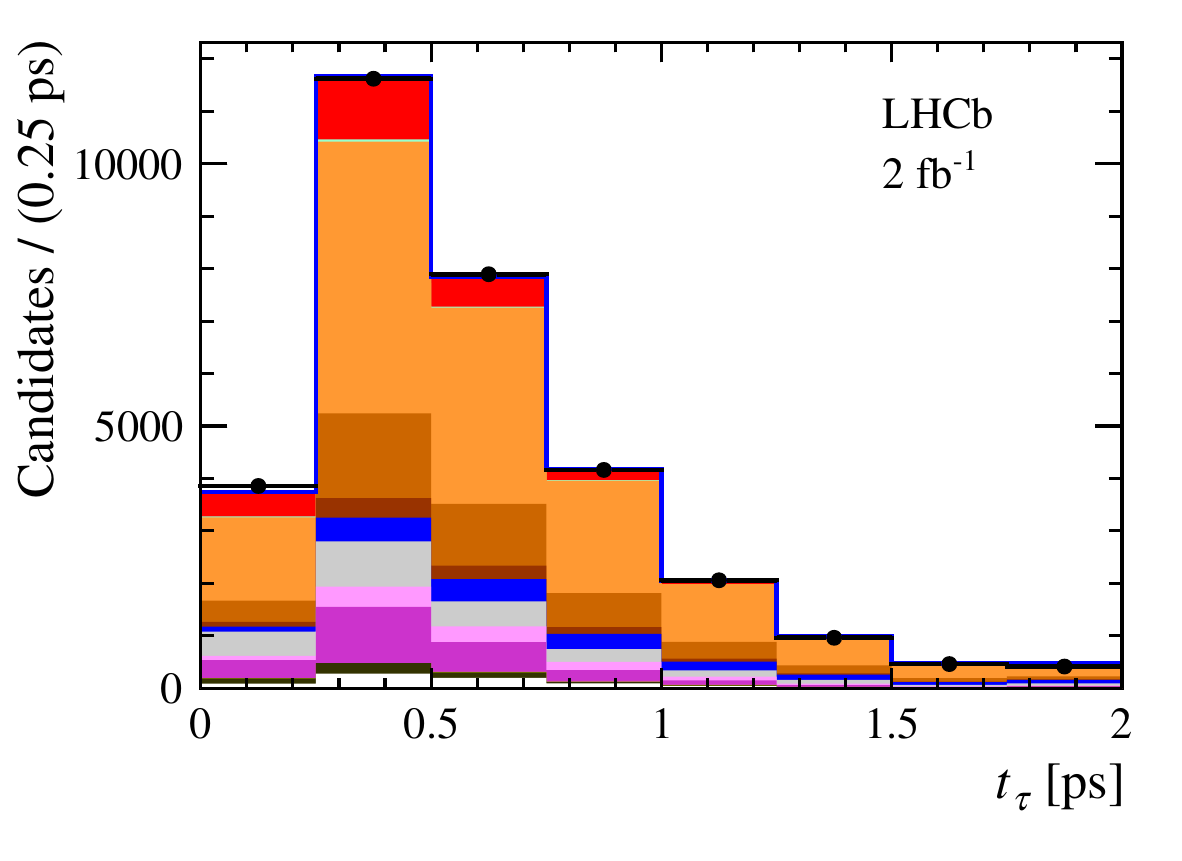}
\end{minipage}
\begin{minipage}[t]{.495\textwidth}
    \vspace{27pt}
    \hspace{23pt}
    \includegraphics[width=0.95\linewidth]{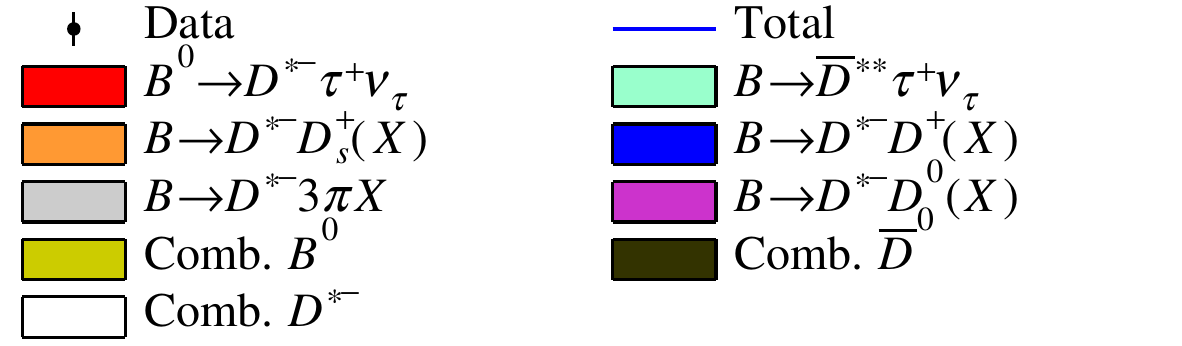}
\end{minipage}

   \caption{Distributions of the fit variables in the \signalmode data sample with the fit result overlaid.}
  \label{fig:fitdataplot}
\end{figure}

The parameters $N_{\text{sig}}$, $N_{\Dsp}$, ${f_{D^+}}$ and ${f_{D^{0}}^{v_1v_2}}$ vary freely in the fit. The fit results are summarized in \cref{tab:fitdata} and the distributions of the fit variables are shown in \cref{fig:fitdataplot}. The fit is performed in two iterations: First, the fractions of $\Dz$ are varied freely and the six $D_s^+$ decay modes are Gaussian constrained, and then a second fit is performed by fixing these to their best fit values. This is the same strategy followed in Refs.~\cite{LHCb-PAPER-2017-017,LHCb-PAPER-2017-027} to determine the statistical uncertainty on the \sig yield. Thus the relative statistical precision on the yield changes from 6.2\% to 5.9\%. The quadratic difference between the statistical uncertainties in the two iterations is treated as a systematic uncertainty from the double-charm decay models. The number of signal events is determined to be $2573\pm156$, where the uncertainty is statistical only. The fit quality is excellent with a $\chi^2$ per degree of freedom of 1.0. 
From studies using pseudoexperiments, the fit is found to be unbiased.

\begin{table}[tbp]
\centering
\caption{Fit results from the three-dimensional signal-extraction fit to $q^2$, $t_{\tau}$ and anti-\Dsp BDT output in the data.}
\begin{tabular}{l
r@{\:$\pm$\:}lr@{\:$\pm$\:}l
r@{\:$\pm$\:}lr@{\:$\pm$\:}l
}
Parameter & \multicolumn{2}{c}{Fit result} & \multicolumn{2}{c}{Constraint}\\ \hline
& \multicolumn{2}{c}{Free} & \multicolumn{2}{c}{} \\ \hline
$ N_{\mathrm{sig}}$ & $2573 $ & $ 156$ & \multicolumn{2}{c}{}\\
$ N_{\Dsp} $ & $20139 $ & $ 508$ & \multicolumn{2}{c}{}\\
$ f_{\Dp} $ & $0.08 $ & $ 0.02$ & \multicolumn{2}{c}{}\\
$f_{D^{0}}^{v_{1}v_{2}}$ & $2.22 $ & $ 0.35$ & \multicolumn{2}{c}{}\\
\hline
& \multicolumn{2}{l}{Constrained} & \multicolumn{2}{c}{}\\ \hline
$ N_{B \rightarrow D^{*-} 3\pi X} $ & $2293 $ & $ 177$ & $2051 $ & $ 200$  \\
$ f_{\Bs \rightarrow \Dstarm \Dsp (X)}$ & $0.13 $ & $ 0.03$ & $0.11 $ & $ 0.04$\\
$f_{D_{s1}^{+}}$ & $0.36 $ & $ 0.03$ & $0.40 $ & $ 0.07$\\
$ f_{\Dsp} $ & $0.60 $ & $ 0.02$ & $0.55 $ & $ 0.03$\\
$f_{D_{s0}^{*+}}$ & $0.06 $ & $ 0.03$ & $0.11 $ & $ 0.04$\\
$f_{\Db^{**}\Dsp (X)}$ & $0.61 $ & $ 0.07$ & $0.34 $ & $ 0.12$ \\
\hline
& \multicolumn{1}{r@{\phantom{${}\pm{}$}}}{Fixed} & \multicolumn{3}{c}{}\\ \hline
$ N_{B_{1}B_{2}} $ & \multicolumn{1}{r@{\phantom{${}\pm{}$}}}{$46$} & \multicolumn{3}{c}{}\\
$ N_{D^{0}}^{\mathrm{same}} $ & \multicolumn{1}{r@{\phantom{${}\pm{}$}}}{$1051$} & \multicolumn{3}{c}{}\\
$N_{\text{fake } \Dzb}$ & \multicolumn{1}{r@{\phantom{${}\pm{}$}}}{$468$} & \multicolumn{3}{c}{}\\
$N_{\text{fake } \Dstarm}$ & \multicolumn{1}{r@{\phantom{${}\pm{}$}}}{$714$} & \multicolumn{3}{c}{}\\
$f_{\Db^{**} \taup \nu}$ & \multicolumn{1}{r@{\phantom{${}\pm{}$}}}{$0.035$} & \multicolumn{3}{c}{}\\
$f_{\taup \to 3\pi \neutb}$ & \multicolumn{1}{r@{\phantom{${}\pm{}$}}}{$0.777$} & \multicolumn{3}{c}{}\\
\hline
\end{tabular}
\label{tab:fitdata}
\end{table}

\subsection{\texorpdfstring{$\boldsymbol{\norm}$}{B0->D*-3pi+/-} yield}

\begin{figure}[t]
  \centering
  \includegraphics[width=.5\linewidth]{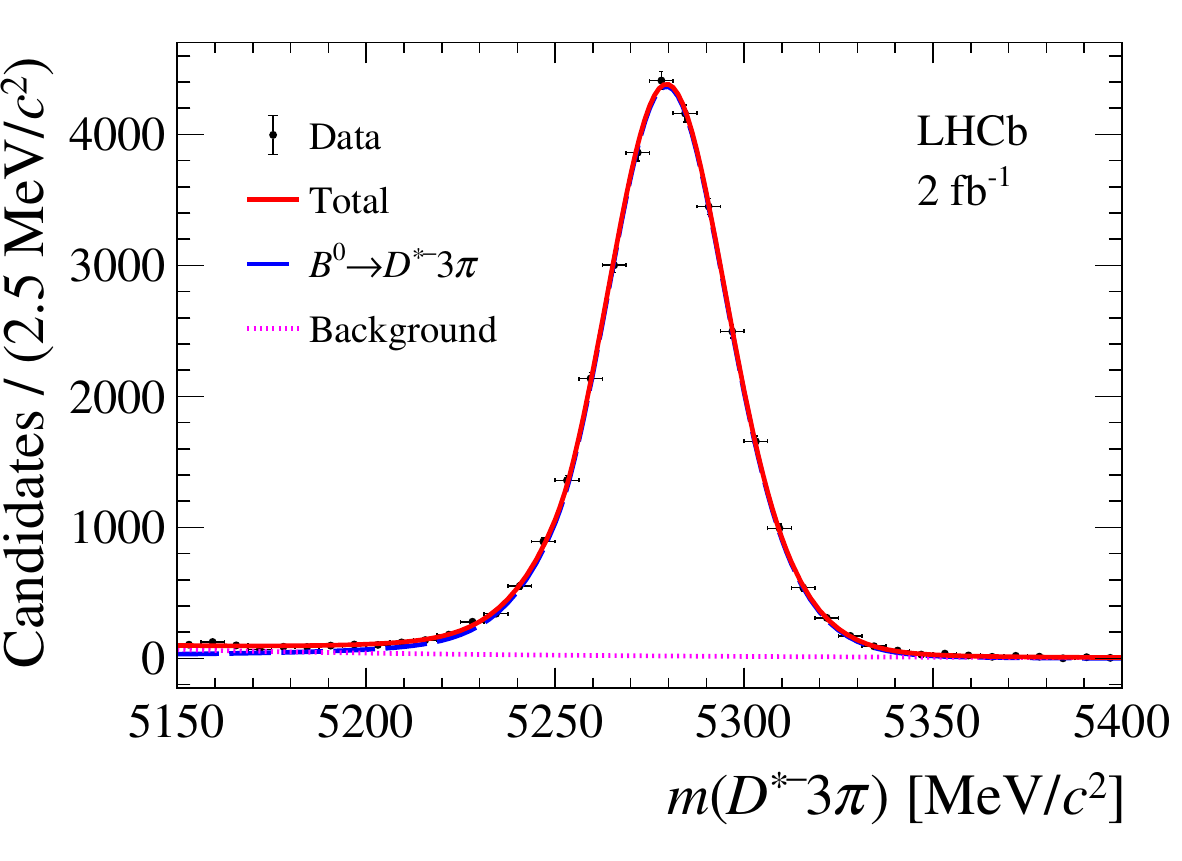}~
  \includegraphics[width=.5\linewidth]{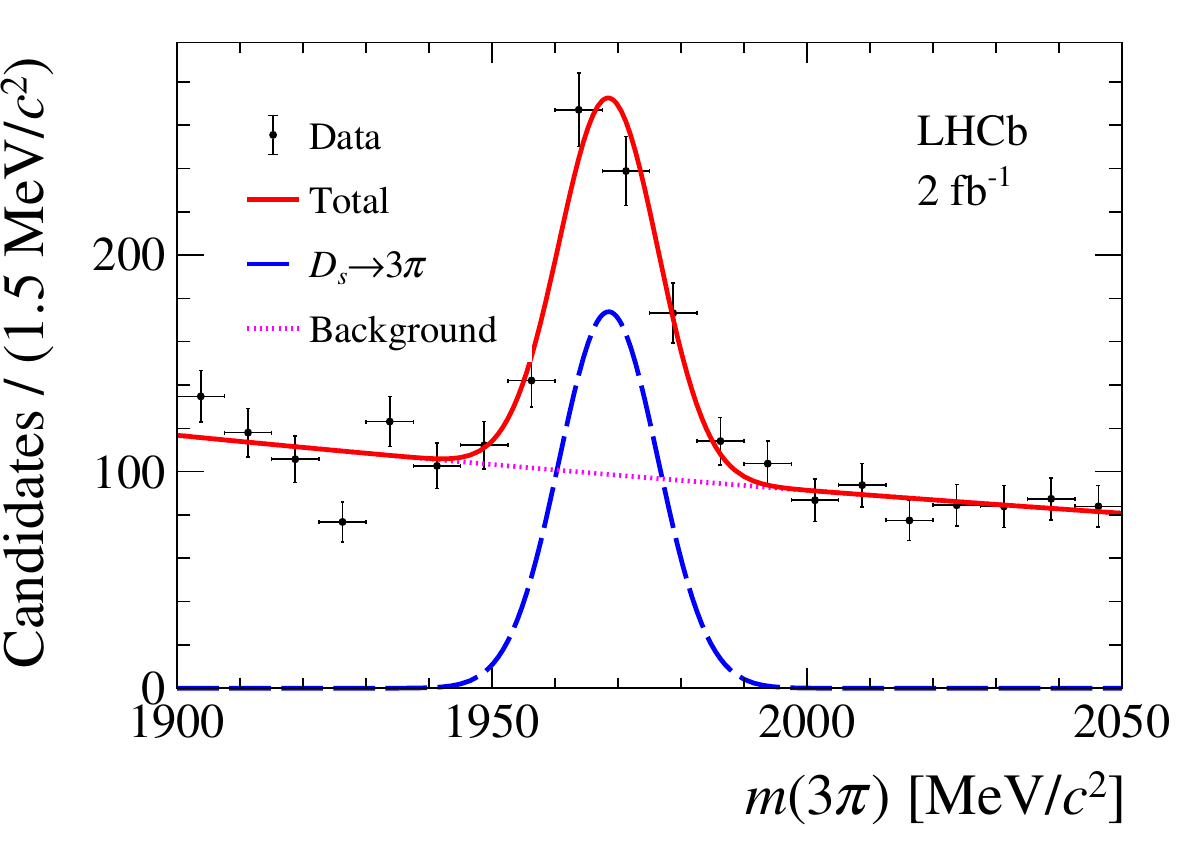}
  
  \caption{Invariant-mass distribution of the (left) $\Dstarm3\pi$ and (right) $3\pi$ system for \norm candidates in data with corresponding fit models superimposed.}
  \label{fig:normfit}
\end{figure}

The \norm yield is estimated from an unbinned maximum-likelihood fit to the $\Dstarm 3\pi^{\pm}$ mass distribution. The signal model consists of a Crystal Ball~(CB) function~\cite{Skwarnicki:1986xj} and two Gaussian functions that share a common mean. The CB shape parameters, the width of the wider Gaussian and the relative proportion of the two Gaussian functions are fixed to the values obtained from simulation. The background component is described by an exponential function. The $m(\Dstarm 3\pi)$ distribution is shown in \cref{fig:normfit} (left) with the fit projection overlaid. The data sample contains \mbox{$B^0 \to \Dstarm \Ds (\to 3\pi)$} decays, which must be subtracted from the \norm yield (\cref{fig:normfit}, right). A fit to the $m(3\pi)$ distribution in the mass range 1800$-$$2100\mevcc$ yields $457\pm37$ decays.  After the subtraction, there are $30800\pm179$ \norm decays. Here, the uncertainties are statistical only.

 The \sig and \norm yields and their efficiencies are used to determine \kappadstar,  yielding a  result of $1.79 \pm 0.11 \pm 0.11$, where the uncertainties are statistical and systematic, respectively. 

\section{SYSTEMATIC UNCERTAINTIES}\label{Sec:syst}
\label{sec:systematics}
Systematic uncertainties on \kappadstar due to the signal and background modeling, selection criteria on the \sig and the \norm decay modes, empty bins in the fit and limited size of simulation samples are considered. 

The simulated \sig decays are weighted with form factors using the Caprini-Lellouch-Neubert parameterization~\cite{Caprini:1997mu} and these are used to produce signal templates for the fit. The systematic uncertainty due to the limited knowledge of the form factors is estimated by changing the baseline to the Boyd-Grinstein-Lebed parameterization~\cite{Boyd:1995sq} and this causes a 1.8\% relative change in the signal yield and a 0.9\% deviation in the signal efficiency. The fraction between the two signal decay modes $f_{\taup \to 3\pi \neutb}$ is fixed in the signal fit. The uncertainty on this is estimated by propagating the uncertainties of the branching fractions and efficiencies. A deviation of 0.3\% is found from alternate fits with $f_{\taup \to 3\pi \neutb}$ varied by $\pm 1 \sigma$. 
Other $\taup$ decays may contribute to the signal, especially those with three charged hadrons in the final state, \eg,\  \Kp\pim\pip,
\Kp\Km\pip or $3\pi\piz\piz$. They are found to have small contributions in the signal sample after full selection. A detailed study with dedicated simulation samples in Refs.~\cite{LHCb-PAPER-2017-017,LHCb-PAPER-2017-027} found a systematic effect of 1.0\% and this is taken as the systematic uncertainty from this source. The fraction of \mbox{$B\to\Db^{**}\taup\neut$} decays, $f_{\Db^{**} \tau \nu}$, is estimated from simulation after considering theoretical assumptions~\cite{bernlochner2021semitauonic}. This fraction is varied by $\pm 50\%$ and a systematic deviation of $^{+1.8}_{-1.9}\%$ is obtained. The \norm fit model depends on  parameters that are fixed to the values obtained in simulation. These are varied by $\pm 1 \sigma$ and the corresponding deviation of 1.0\% is assigned as a systematic uncertainty.

The modeling of double-charm, prompt and combinatorial backgrounds contribute to the systematic uncertainty on \kappadstar. The \Ds decays in simulation are corrected by the weights obtained in the fit to the \mbox{$\Ds \to 3\pi X $} control sample. These weights are varied by their uncertainties after accounting for correlations and alternate signal-extraction fits are performed with the new \Ds templates. This results in a 1.0\% systematic deviation in \kappadstar. The fixed fraction of \mbox{$\Ds \to R a_1$} in the fit to the \mbox{$\Ds \to 3\pi X $} control sample is varied by 30\% and this contributes to a 1.5\% systematic deviation. This variation corresponds to a change of 1 in the $\chi^2$ of the signal-extraction fit.

The variables in the signal-extraction fit, $q^2$, $t_{\tau}$ and the anti-\Dsp BDT output, can have a non-negligible correlation with some important kinematic variables like $m(\Dstarm3\pi)$, $m(3\pi)$, $\text{min}[m(\pip \pim)]$ and $\text{max}[m(\pip \pim)]$ that are used in background studies. The background template shapes that are derived from simulation are varied with weights that are functions of these kinematic variables. The signal-extraction fit is repeated with the alternate templates for the different background categories to estimate the systematic effect on the results. The variations in the \Ds, \Dz, \Dp and prompt template shapes result in 0.3\%, 1.2\%, $^{+2.2}_{-0.8}\%$ and 1.2\% deviations, respectively, to \kappadstar. The effect of fixing the \Dp and \Dz fractions ($f_{\Dp}$ and $f_{D^{0}}^{v_{1}v_{2}}$) as well as combinatorial yields are estimated by varying the fixed values by their uncertainties.

The efficiencies of particle identification requirements are estimated using calibration samples.  These samples are used to parametrize the efficiency in bins of the kinematic variables of a given track such as $p$, \pt and hits in the tracking system. The choice of the binning scheme for this parametrization contributes a 0.5\% systematic uncertainty. The effect of the finite size of the calibration sample is negligible. The systematic effects of the correction weights applied to the simulation are obtained either by varying the correction factor or removing the correction altogether. Changes in kinematic weighting, $3\pi$ vertex error correction and $3\pi$ model correction in \norm decays result in 0.7\%, 0.9\% and 1.0\% systematic deviations, respectively, on \kappadstar. The difference in data and simulation at the preselection level affects the efficiency determination, contributing 2.0\% to the systematic uncertainty. The statistical uncertainty on the \sig and \norm efficiencies due to the limited size of the simulation sample also contribute to the systematic uncertainty.

The three-dimensional templates used in the fit include empty bins. The systematic deviation due to these empty bins is estimated by building a three-dimensional density function using kernel density estimators. These are built for all the components and then transformed into a histogram template and the fit is repeated with the new templates. This contributes to a 1.3\% relative systematic deviation. 

The template PDFs used in the signal fit are mostly derived from simulation. Therefore, the size of these simulation samples has a major impact on the results. Alternative templates obtained after resampling the default templates via a bootstrap procedure are used to repeat the fit 1000 times and the resulting deviation of 2.0\% is taken as the systematic uncertainty from this source. The large simulation samples produced using the \textsc{redecay}~\cite{LHCb-DP-2018-004} technique helped in reducing this systematic uncertainty to half the value obtained in Refs.~\cite{LHCb-PAPER-2017-017,LHCb-PAPER-2017-027}. A summary of all the systematic uncertainties on \kappadstar is given in \cref{tab:systsum}. The total systematic uncertainty is obtained by adding the individual contributions in quadrature.

\begin{table}[tbp]
\centering
\caption{Summary of relative systematic uncertainties on the ratio \kappadstar.}
\begin{tabular}{lc}
\hline
Source &  Systematic uncertainty  (\%)  \\
 \hline
 Signal decay template shape&   1.8 \\
Signal decay efficiency&   0.9 \\
Fractions of signal $\taup$ decays & $0.3$\\
Possible contributions from other $\taup$ decays & 1.0  \\
Fixing the $\Db^{**}\taup \neut$ and $D_s^{**+} \taup \neut$ fractions & $^{+1.8}_{-1.9}$ \\
Normalization mode PDF choice & 1.0 \\
Knowledge of the $D_s^+ \to 3\pi X$ decay model & 1.0\\
~~~~Specifically the $\Dsp \to a_1 X$ fraction & $1.5$ \\
$B\to D^{*-}D_s^+(X)$ template shapes& 0.3\\
$B\to D^{*-}D^0(X)$ template shapes&  1.2 \\
$B \to D^{*-}D^+(X)$ template shapes & $^{+2.2}_{-0.8}$\\
Fixing $B\to \Dstarm \Ds (X)$ bkg model parameters & $1.1$ \\
Fixing $B\to D^{*-}D^0 (X)$ bkg model parameters & $1.5$ \\
$B \to D^{*-}3\pi X$ template shapes & 1.2\\
Combinatorial background normalization& $^{+0.5}_{-0.6}$\\
PID efficiency&  0.5\\
Kinematic reweighting & 0.7 \\
Vertex error correction & 0.9 \\
Normalization mode efficiency~(modeling of $m(3\pi)$)&  1.0 \\
Preselection efficiency & 2.0 \\
Signal efficiency~(size of simulation sample) & 1.1 \\
Normalization efficiency (size of simulation sample) & 1.1 \\
Empty bins in templates& 1.3 \\ 
PDF shapes uncertainty (size of simulation sample) &  2.0 \\
\hline
Total systematic uncertainty & $^{+6.2}_{-5.9}$  \\
\hline
Total statistical uncertainty & 5.9  \\
\hline
\end{tabular}
\label{tab:systsum}
\end{table}

\section{CONCLUSION}
\label{sec:conclusion}

In conclusion, using $pp$ collision data collected in 2015 and 2016 by the LHCb experiment and corresponding to an integrated luminosity of 2~\invfb, the ratio of the branching fractions of \sig and \norm decays is measured as
\begin{equation}
\kappadstar = 1.79 \pm 0.11(\text{stat}) \pm 0.11(\text{syst}) \,. \nonumber
\end{equation}
The result is in good agreement with the previous LHCb measurement~\cite{LHCb-PAPER-2017-017,LHCb-PAPER-2017-027}. 
The improved analysis procedure results in an increase in signal efficiency and a decrease of the relative systematic uncertainty from 9\% to 6\%.
Using the most recent branching fraction measurements \mbox{$\BF(\Bz\to\Dstarm 3\pion) = (7.21 \pm 0.29) \times 10^{-3}$} and \mbox{$\BF(\Bz\to \Dstarm \mup \neum) = (4.97 \pm 0.12)\%$}~\cite{PDG2022}, the branching fraction
\begin{equation}
\BF( \Bz\to\Dstarm\taup\neut)=(1.29\pm 0.08\stat \pm0.08\syst \pm 0.05~(\text{ext})) \times 10^{-2}\,, \nonumber
\end{equation}
and the ratio of branching fractions of \sig and $\decay{\Bd}{\Dstarm \mup \neum}$
\begin{equation}
\rdstarm=0.260\pm 0.015\stat \pm 0.016\syst \pm 0.012~(\text{ext})\,, \nonumber
\end{equation}
are obtained, where the third uncertainties are due to the uncertainties on the external branching fractions.
This result is compatible with the present world average and with the SM expectation ($0.254 \pm 0.005$)~\cite{hflav_2021}. When combined with the previous results~\cite{LHCb-PAPER-2017-017,LHCb-PAPER-2017-027}, the values of \kappadstar and \rdstarm are
\begin{align}
\kappadstar_{\rm comb} & =  1.84 \pm  0.08\stat \pm 0.10\syst\, \text{and}
\nonumber \\
\rdstarm_{\rm comb}  & =  0.267\pm 0.012\stat \pm 0.015\syst \pm0.013~(\text{ext}) \,.  \nonumber
\end{align}
The combined \rdstarm is obtained from the \kappadstar combination and the branching fractions of the normalization channels. This combination leads to one of the most precise measurements of \rdstar to date.
\clearpage

\section*{ACKNOWLEDGMENTS}
%
%
\noindent We express our gratitude to our colleagues in the CERN
accelerator departments for the excellent performance of the LHC. We
thank the technical and administrative staff at the LHCb
institutes.
We acknowledge support from CERN and from the national agencies:
CAPES, CNPq, FAPERJ and FINEP (Brazil); 
MOST and NSFC (China); 
CNRS/IN2P3 (France); 
BMBF, DFG and MPG (Germany); 
INFN (Italy); 
NWO (Netherlands); 
MNiSW and NCN (Poland); 
MEN/IFA (Romania); 
MICINN (Spain); 
SNSF and SER (Switzerland); 
NASU (Ukraine); 
STFC (United Kingdom); 
DOE NP and NSF (USA).
We acknowledge the computing resources that are provided by CERN, IN2P3
(France), KIT and DESY (Germany), INFN (Italy), SURF (Netherlands),
PIC (Spain), GridPP (United Kingdom), 
CSCS (Switzerland), IFIN-HH (Romania), CBPF (Brazil),
Polish WLCG  (Poland) and NERSC (USA).
We are indebted to the communities behind the multiple open-source
software packages on which we depend.
Individual groups or members have received support from
ARC and ARDC (Australia);
Minciencias (Colombia);
AvH Foundation (Germany);
EPLANET, Marie Sk\l{}odowska-Curie Actions and ERC (European Union);
A*MIDEX, ANR, IPhU and Labex P2IO, and R\'{e}gion Auvergne-Rh\^{o}ne-Alpes (France);
Key Research Program of Frontier Sciences of CAS, CAS PIFI, CAS CCEPP, 
Fundamental Research Funds for the Central Universities, 
and Science and Technology Program of Guangzhou (China);
GVA, XuntaGal, GENCAT and Prog.~Atracci\'on Talento, CM (Spain);
SRC (Sweden);
the Leverhulme Trust, the Royal Society
 and UKRI (United Kingdom).




\newpage
\addcontentsline{toc}{section}{References}
\bibliographystyle{LHCb}
\bibliography{main,standard,LHCb-PAPER,LHCb-CONF,LHCb-DP,LHCb-TDR,exp,pheno}

\ifx\mcitethebibliography\mciteundefinedmacro
\PackageError{LHCb.bst}{mciteplus.sty has not been loaded}
{This bibstyle requires the use of the mciteplus package.}\fi
\providecommand{\href}[2]{#2}
\begin{mcitethebibliography}{10}
\mciteSetBstSublistMode{n}
\mciteSetBstMaxWidthForm{subitem}{\alph{mcitesubitemcount})}
\mciteSetBstSublistLabelBeginEnd{\mcitemaxwidthsubitemform\space}
{\relax}{\relax}

\bibitem{Lees_2012}
BaBar collaboration, J.~P. Lees {\em et~al.},
  \ifthenelse{\boolean{articletitles}}{\emph{{Evidence for an excess of
  $\bar{B} \to D^{(*)} \tau^-\bar{\nu}_\tau$ decays}},
  }{}\href{https://doi.org/10.1103/PhysRevLett.109.101802}{Phys.\ Rev.\ Lett.\
  \textbf{109} (2012) 101802},
  \href{http://arxiv.org/abs/1205.5442}{{\normalfont\ttfamily
  arXiv:1205.5442}}\relax
\mciteBstWouldAddEndPuncttrue
\mciteSetBstMidEndSepPunct{\mcitedefaultmidpunct}
{\mcitedefaultendpunct}{\mcitedefaultseppunct}\relax
\EndOfBibitem
\bibitem{belle_collaboration_measurement_2015}
Belle collaboration, M.~Huschle {\em et~al.},
  \ifthenelse{\boolean{articletitles}}{\emph{{Measurement of the branching
  ratio of $\bar{B} \to D^{(\ast)} \tau^- \bar{\nu}_\tau$ relative to $\bar{B}
  \to D^{(\ast)} \ell^- \bar{\nu}_\ell$ decays with hadronic tagging at
  Belle}}, }{}\href{https://doi.org/10.1103/PhysRevD.92.072014}{Phys.\ Rev.\
  \textbf{D92} (2015) 072014},
  \href{http://arxiv.org/abs/1507.03233}{{\normalfont\ttfamily
  arXiv:1507.03233}}\relax
\mciteBstWouldAddEndPuncttrue
\mciteSetBstMidEndSepPunct{\mcitedefaultmidpunct}
{\mcitedefaultendpunct}{\mcitedefaultseppunct}\relax
\EndOfBibitem
\bibitem{the_belle_collaboration_measurement_2018}
Belle collaboration, S.~Hirose {\em et~al.},
  \ifthenelse{\boolean{articletitles}}{\emph{{Measurement of the $\tau$ lepton
  polarization and $R(D^*)$ in the decay $\bar{B} \rightarrow D^* \tau^-
  \bar{\nu}_\tau$ with one-prong hadronic $\tau$ decays at Belle}},
  }{}\href{https://doi.org/10.1103/PhysRevD.97.012004}{Phys.\ Rev.\
  \textbf{D97} (2018) 012004},
  \href{http://arxiv.org/abs/1709.00129}{{\normalfont\ttfamily
  arXiv:1709.00129}}\relax
\mciteBstWouldAddEndPuncttrue
\mciteSetBstMidEndSepPunct{\mcitedefaultmidpunct}
{\mcitedefaultendpunct}{\mcitedefaultseppunct}\relax
\EndOfBibitem
\bibitem{LHCb-PAPER-2017-017}
LHCb collaboration, R.~Aaij {\em et~al.},
  \ifthenelse{\boolean{articletitles}}{\emph{{Measurement of the ratio of the
  \mbox{$\mathcal{B}(\decay{\Bz}{D^{\ast-} \taup \nu_{\tau}})$} and
  \mbox{$\mathcal{B}(\decay{\Bz}{D^{\ast-}\mup\nu_{\mu}})$} branching fractions
  using three-prong $\tau$-lepton decays}},
  }{}\href{https://doi.org/10.1103/PhysRevLett.120.171802}{Phys.\ Rev.\ Lett.\
  \textbf{120} (2018) 171802},
  \href{http://arxiv.org/abs/1708.08856}{{\normalfont\ttfamily
  arXiv:1708.08856}}\relax
\mciteBstWouldAddEndPuncttrue
\mciteSetBstMidEndSepPunct{\mcitedefaultmidpunct}
{\mcitedefaultendpunct}{\mcitedefaultseppunct}\relax
\EndOfBibitem
\bibitem{LHCb-PAPER-2017-027}
LHCb collaboration, R.~Aaij {\em et~al.},
  \ifthenelse{\boolean{articletitles}}{\emph{{Test of lepton flavor
  universality by the measurement of the \mbox{\decay{\Bz}{D^{\ast-} \taup
  \nu_{\tau}}} branching fraction using three-prong $\tau$ decays}},
  }{}\href{https://doi.org/10.1103/PhysRevD.97.072013}{Phys.\ Rev.\
  \textbf{D97} (2018) 072013},
  \href{http://arxiv.org/abs/1711.02505}{{\normalfont\ttfamily
  arXiv:1711.02505}}\relax
\mciteBstWouldAddEndPuncttrue
\mciteSetBstMidEndSepPunct{\mcitedefaultmidpunct}
{\mcitedefaultendpunct}{\mcitedefaultseppunct}\relax
\EndOfBibitem
\bibitem{LHCb-PAPER-2022-039}
LHCb collaboration, R.~Aaij {\em et~al.},
  \ifthenelse{\boolean{articletitles}}{\emph{{Measurement of the ratio of
  branching fractions $\mathcal{R}(\Dstar)$ and $\mathcal{R}(\Dz)$ }},
  }{}\href{http://arxiv.org/abs/2302.02886}{{\normalfont\ttfamily
  arXiv:2302.02886}}, {submitted to Phys.~Rev.~Lett.}\relax
\mciteBstWouldAddEndPunctfalse
\mciteSetBstMidEndSepPunct{\mcitedefaultmidpunct}
{}{\mcitedefaultseppunct}\relax
\EndOfBibitem
\bibitem{hflav_2021}
HFLAV collaboration, Y.~Amhis {\em et~al.},
  \ifthenelse{\boolean{articletitles}}{\emph{{Averages of $b$-hadron,
  $c$-hadron, and $\tau$-lepton properties as of 2021}},
  }{}\href{http://arxiv.org/abs/2206.07501}{{\normalfont\ttfamily
  arXiv:2206.07501}}\relax
\mciteBstWouldAddEndPuncttrue
\mciteSetBstMidEndSepPunct{\mcitedefaultmidpunct}
{\mcitedefaultendpunct}{\mcitedefaultseppunct}\relax
\EndOfBibitem
\bibitem{Fajfer_2012}
S.~Fajfer, J.~F. Kamenik, and I.~Nisandzic,
  \ifthenelse{\boolean{articletitles}}{\emph{{On the $B \to D^* \tau \bar
  \nu_{\tau}$ sensitivity to new physics}},
  }{}\href{https://doi.org/10.1103/PhysRevD.85.094025}{Phys.\ Rev.\
  \textbf{D85} (2012) 094025},
  \href{http://arxiv.org/abs/1203.2654}{{\normalfont\ttfamily
  arXiv:1203.2654}}\relax
\mciteBstWouldAddEndPuncttrue
\mciteSetBstMidEndSepPunct{\mcitedefaultmidpunct}
{\mcitedefaultendpunct}{\mcitedefaultseppunct}\relax
\EndOfBibitem
\bibitem{Fajfer_2016}
S.~Fajfer and N.~Košnik, \ifthenelse{\boolean{articletitles}}{\emph{Vector
  leptoquark resolution of {$R_{K}$} and {$R_{D^{(\ast)}}$} puzzles},
  }{}\href{https://doi.org/10.1016/j.physletb.2016.02.018}{Physics Letters B
  \textbf{755} (2016) 270–274},
  \href{http://arxiv.org/abs/1511.06024}{{\normalfont\ttfamily
  arXiv:1511.06024}}\relax
\mciteBstWouldAddEndPuncttrue
\mciteSetBstMidEndSepPunct{\mcitedefaultmidpunct}
{\mcitedefaultendpunct}{\mcitedefaultseppunct}\relax
\EndOfBibitem
\bibitem{crivellin_simultaneous_2017}
A.~Crivellin, D.~Müller, and T.~Ota,
  \ifthenelse{\boolean{articletitles}}{\emph{Simultaneous {explanation} of
  ${R}({D}^{(*)})$ and $b\to s\mu^+\mu^-$: {the} {last} {scalar} {leptoquarks}
  {standing}}, }{}\href{https://doi.org/10.1007/JHEP09(2017)040}{JHEP
  \textbf{2017} (2017) 40},
  \href{http://arxiv.org/abs/1703.09226}{{\normalfont\ttfamily
  arXiv:1703.09226}}\relax
\mciteBstWouldAddEndPuncttrue
\mciteSetBstMidEndSepPunct{\mcitedefaultmidpunct}
{\mcitedefaultendpunct}{\mcitedefaultseppunct}\relax
\EndOfBibitem
\bibitem{PDG2022}
Particle Data Group, R.~L. Workman {\em et~al.},
  \ifthenelse{\boolean{articletitles}}{\emph{{\href{http://pdg.lbl.gov/}{Review
  of particle physics}}}, }{}\href{https://doi.org/10.1093/ptep/ptac097}{Prog.\
  Theor.\ Exp.\ Phys.\  \textbf{2022} (2022) 083C01}\relax
\mciteBstWouldAddEndPuncttrue
\mciteSetBstMidEndSepPunct{\mcitedefaultmidpunct}
{\mcitedefaultendpunct}{\mcitedefaultseppunct}\relax
\EndOfBibitem
\bibitem{LHCb-DP-2008-001}
LHCb collaboration, A.~A. Alves~Jr.\ {\em et~al.},
  \ifthenelse{\boolean{articletitles}}{\emph{{The \lhcb detector at the LHC}},
  }{}\href{https://doi.org/10.1088/1748-0221/3/08/S08005}{JINST \textbf{3}
  (2008) S08005}\relax
\mciteBstWouldAddEndPuncttrue
\mciteSetBstMidEndSepPunct{\mcitedefaultmidpunct}
{\mcitedefaultendpunct}{\mcitedefaultseppunct}\relax
\EndOfBibitem
\bibitem{LHCb-DP-2014-002}
LHCb collaboration, R.~Aaij {\em et~al.},
  \ifthenelse{\boolean{articletitles}}{\emph{{LHCb detector performance}},
  }{}\href{https://doi.org/10.1142/S0217751X15300227}{Int.\ J.\ Mod.\ Phys.\
  \textbf{A30} (2015) 1530022},
  \href{http://arxiv.org/abs/1412.6352}{{\normalfont\ttfamily
  arXiv:1412.6352}}\relax
\mciteBstWouldAddEndPuncttrue
\mciteSetBstMidEndSepPunct{\mcitedefaultmidpunct}
{\mcitedefaultendpunct}{\mcitedefaultseppunct}\relax
\EndOfBibitem
\bibitem{LHCb-DP-2014-001}
R.~Aaij {\em et~al.}, \ifthenelse{\boolean{articletitles}}{\emph{{Performance
  of the LHCb Vertex Locator}},
  }{}\href{https://doi.org/10.1088/1748-0221/9/09/P09007}{JINST \textbf{9}
  (2014) P09007}, \href{http://arxiv.org/abs/1405.7808}{{\normalfont\ttfamily
  arXiv:1405.7808}}\relax
\mciteBstWouldAddEndPuncttrue
\mciteSetBstMidEndSepPunct{\mcitedefaultmidpunct}
{\mcitedefaultendpunct}{\mcitedefaultseppunct}\relax
\EndOfBibitem
\bibitem{LHCb-DP-2013-003}
R.~Arink {\em et~al.}, \ifthenelse{\boolean{articletitles}}{\emph{{Performance
  of the LHCb Outer Tracker}},
  }{}\href{https://doi.org/10.1088/1748-0221/9/01/P01002}{JINST \textbf{9}
  (2014) P01002}, \href{http://arxiv.org/abs/1311.3893}{{\normalfont\ttfamily
  arXiv:1311.3893}}\relax
\mciteBstWouldAddEndPuncttrue
\mciteSetBstMidEndSepPunct{\mcitedefaultmidpunct}
{\mcitedefaultendpunct}{\mcitedefaultseppunct}\relax
\EndOfBibitem
\bibitem{LHCb-DP-2017-001}
P.~d'Argent {\em et~al.}, \ifthenelse{\boolean{articletitles}}{\emph{{Improved
  performance of the LHCb Outer Tracker in LHC Run 2}},
  }{}\href{https://doi.org/10.1088/1748-0221/12/11/P11016}{JINST \textbf{12}
  (2017) P11016}, \href{http://arxiv.org/abs/1708.00819}{{\normalfont\ttfamily
  arXiv:1708.00819}}\relax
\mciteBstWouldAddEndPuncttrue
\mciteSetBstMidEndSepPunct{\mcitedefaultmidpunct}
{\mcitedefaultendpunct}{\mcitedefaultseppunct}\relax
\EndOfBibitem
\bibitem{LHCb-DP-2012-003}
M.~Adinolfi {\em et~al.},
  \ifthenelse{\boolean{articletitles}}{\emph{{Performance of the \lhcb RICH
  detector at the LHC}},
  }{}\href{https://doi.org/10.1140/epjc/s10052-013-2431-9}{Eur.\ Phys.\ J.\
  \textbf{C73} (2013) 2431},
  \href{http://arxiv.org/abs/1211.6759}{{\normalfont\ttfamily
  arXiv:1211.6759}}\relax
\mciteBstWouldAddEndPuncttrue
\mciteSetBstMidEndSepPunct{\mcitedefaultmidpunct}
{\mcitedefaultendpunct}{\mcitedefaultseppunct}\relax
\EndOfBibitem
\bibitem{LHCb-DP-2012-002}
A.~A. Alves~Jr.\ {\em et~al.},
  \ifthenelse{\boolean{articletitles}}{\emph{{Performance of the LHCb muon
  system}}, }{}\href{https://doi.org/10.1088/1748-0221/8/02/P02022}{JINST
  \textbf{8} (2013) P02022},
  \href{http://arxiv.org/abs/1211.1346}{{\normalfont\ttfamily
  arXiv:1211.1346}}\relax
\mciteBstWouldAddEndPuncttrue
\mciteSetBstMidEndSepPunct{\mcitedefaultmidpunct}
{\mcitedefaultendpunct}{\mcitedefaultseppunct}\relax
\EndOfBibitem
\bibitem{LHCb-DP-2012-004}
R.~Aaij {\em et~al.}, \ifthenelse{\boolean{articletitles}}{\emph{{The \lhcb
  trigger and its performance in 2011}},
  }{}\href{https://doi.org/10.1088/1748-0221/8/04/P04022}{JINST \textbf{8}
  (2013) P04022}, \href{http://arxiv.org/abs/1211.3055}{{\normalfont\ttfamily
  arXiv:1211.3055}}\relax
\mciteBstWouldAddEndPuncttrue
\mciteSetBstMidEndSepPunct{\mcitedefaultmidpunct}
{\mcitedefaultendpunct}{\mcitedefaultseppunct}\relax
\EndOfBibitem
\bibitem{BBDT}
V.~V. Gligorov and M.~Williams,
  \ifthenelse{\boolean{articletitles}}{\emph{{Efficient, reliable and fast
  high-level triggering using a bonsai boosted decision tree}},
  }{}\href{https://doi.org/10.1088/1748-0221/8/02/P02013}{JINST \textbf{8}
  (2013) P02013}, \href{http://arxiv.org/abs/1210.6861}{{\normalfont\ttfamily
  arXiv:1210.6861}}\relax
\mciteBstWouldAddEndPuncttrue
\mciteSetBstMidEndSepPunct{\mcitedefaultmidpunct}
{\mcitedefaultendpunct}{\mcitedefaultseppunct}\relax
\EndOfBibitem
\bibitem{LHCb-PROC-2015-018}
T.~Likhomanenko {\em et~al.}, \ifthenelse{\boolean{articletitles}}{\emph{{LHCb
  topological trigger reoptimization}},
  }{}\href{https://doi.org/10.1088/1742-6596/664/8/082025}{J.\ Phys.\ Conf.\
  Ser.\  \textbf{664} (2015) 082025},
  \href{http://arxiv.org/abs/1510.00572}{{\normalfont\ttfamily
  arXiv:1510.00572}}\relax
\mciteBstWouldAddEndPuncttrue
\mciteSetBstMidEndSepPunct{\mcitedefaultmidpunct}
{\mcitedefaultendpunct}{\mcitedefaultseppunct}\relax
\EndOfBibitem
\bibitem{Sjostrand:2007gs}
T.~Sj\"{o}strand, S.~Mrenna, and P.~Skands,
  \ifthenelse{\boolean{articletitles}}{\emph{{A brief introduction to PYTHIA
  8.1}}, }{}\href{https://doi.org/10.1016/j.cpc.2008.01.036}{Comput.\ Phys.\
  Commun.\  \textbf{178} (2008) 852},
  \href{http://arxiv.org/abs/0710.3820}{{\normalfont\ttfamily
  arXiv:0710.3820}}\relax
\mciteBstWouldAddEndPuncttrue
\mciteSetBstMidEndSepPunct{\mcitedefaultmidpunct}
{\mcitedefaultendpunct}{\mcitedefaultseppunct}\relax
\EndOfBibitem
\bibitem{Sjostrand:2006za}
T.~Sj\"{o}strand, S.~Mrenna, and P.~Skands,
  \ifthenelse{\boolean{articletitles}}{\emph{{PYTHIA 6.4 physics and manual}},
  }{}\href{https://doi.org/10.1088/1126-6708/2006/05/026}{JHEP \textbf{05}
  (2006) 026}, \href{http://arxiv.org/abs/hep-ph/0603175}{{\normalfont\ttfamily
  arXiv:hep-ph/0603175}}\relax
\mciteBstWouldAddEndPuncttrue
\mciteSetBstMidEndSepPunct{\mcitedefaultmidpunct}
{\mcitedefaultendpunct}{\mcitedefaultseppunct}\relax
\EndOfBibitem
\bibitem{LHCb-PROC-2010-056}
I.~Belyaev {\em et~al.}, \ifthenelse{\boolean{articletitles}}{\emph{{Handling
  of the generation of primary events in Gauss, the LHCb simulation
  framework}}, }{}\href{https://doi.org/10.1088/1742-6596/331/3/032047}{J.\
  Phys.\ Conf.\ Ser.\  \textbf{331} (2011) 032047}\relax
\mciteBstWouldAddEndPuncttrue
\mciteSetBstMidEndSepPunct{\mcitedefaultmidpunct}
{\mcitedefaultendpunct}{\mcitedefaultseppunct}\relax
\EndOfBibitem
\bibitem{Lange:2001uf}
D.~J. Lange, \ifthenelse{\boolean{articletitles}}{\emph{{The EvtGen particle
  decay simulation package}},
  }{}\href{https://doi.org/10.1016/S0168-9002(01)00089-4}{Nucl.\ Instrum.\
  Meth.\  \textbf{A462} (2001) 152}\relax
\mciteBstWouldAddEndPuncttrue
\mciteSetBstMidEndSepPunct{\mcitedefaultmidpunct}
{\mcitedefaultendpunct}{\mcitedefaultseppunct}\relax
\EndOfBibitem
\bibitem{davidson2015photos}
N.~Davidson, T.~Przedzinski, and Z.~Was,
  \ifthenelse{\boolean{articletitles}}{\emph{{PHOTOS interface in C++:
  Technical and physics documentation}},
  }{}\href{https://doi.org/https://doi.org/10.1016/j.cpc.2015.09.013}{Comp.\
  Phys.\ Comm.\  \textbf{199} (2016) 86},
  \href{http://arxiv.org/abs/1011.0937}{{\normalfont\ttfamily
  arXiv:1011.0937}}\relax
\mciteBstWouldAddEndPuncttrue
\mciteSetBstMidEndSepPunct{\mcitedefaultmidpunct}
{\mcitedefaultendpunct}{\mcitedefaultseppunct}\relax
\EndOfBibitem
\bibitem{Chiral_lagrangian}
I.~M. Nugent {\em et~al.}, \ifthenelse{\boolean{articletitles}}{\emph{Resonance
  chiral lagrangian currents and experimental data for
  ${\ensuremath{\tau}}^{\mathbf{\ensuremath{-}}}\ensuremath{\rightarrow}{\ensuremath{\pi}}^{\mathbf{\ensuremath{-}}}{\ensuremath{\pi}}^{\mathbf{\ensuremath{-}}}{\ensuremath{\pi}}^{\mathbf{+}}{\ensuremath{\nu}}_{\ensuremath{\tau}}$},
  }{}\href{https://doi.org/10.1103/PhysRevD.88.093012}{Phys.\ Rev.\
  \textbf{D88} (2013) 093012},
  \href{http://arxiv.org/abs/1310.1053}{{\normalfont\ttfamily
  arXiv:1310.1053}}\relax
\mciteBstWouldAddEndPuncttrue
\mciteSetBstMidEndSepPunct{\mcitedefaultmidpunct}
{\mcitedefaultendpunct}{\mcitedefaultseppunct}\relax
\EndOfBibitem
\bibitem{DAVIDSON2012821}
N.~Davidson {\em et~al.}, \ifthenelse{\boolean{articletitles}}{\emph{Universal
  interface of tauola: Technical and physics documentation},
  }{}\href{https://doi.org/https://doi.org/10.1016/j.cpc.2011.12.009}{Computer
  Physics Communications \textbf{183} (2012) 821},
  \href{http://arxiv.org/abs/1002.0543}{{\normalfont\ttfamily
  arXiv:1002.0543}}\relax
\mciteBstWouldAddEndPuncttrue
\mciteSetBstMidEndSepPunct{\mcitedefaultmidpunct}
{\mcitedefaultendpunct}{\mcitedefaultseppunct}\relax
\EndOfBibitem
\bibitem{NUGENT201438}
I.~M. Nugent, \ifthenelse{\boolean{articletitles}}{\emph{Invariant mass spectra
  of $\tau^- \to h^- h^- h^+ \nu_\tau$ decays},
  }{}\href{https://doi.org/https://doi.org/10.1016/j.nuclphysbps.2014.09.010}{Nuclear
  Physics B - Proceedings Supplements \textbf{253-255} (2014) 38},
  \href{http://arxiv.org/abs/1301.7105}{{\normalfont\ttfamily
  arXiv:1301.7105}}, The Twelfth International Workshop on Tau-Lepton Physics
  (TAU2012)\relax
\mciteBstWouldAddEndPuncttrue
\mciteSetBstMidEndSepPunct{\mcitedefaultmidpunct}
{\mcitedefaultendpunct}{\mcitedefaultseppunct}\relax
\EndOfBibitem
\bibitem{Allison:2006ve}
Geant4 collaboration, J.~Allison {\em et~al.},
  \ifthenelse{\boolean{articletitles}}{\emph{{Geant4 developments and
  applications}}, }{}\href{https://doi.org/10.1109/TNS.2006.869826}{IEEE
  Trans.\ Nucl.\ Sci.\  \textbf{53} (2006) 270}\relax
\mciteBstWouldAddEndPuncttrue
\mciteSetBstMidEndSepPunct{\mcitedefaultmidpunct}
{\mcitedefaultendpunct}{\mcitedefaultseppunct}\relax
\EndOfBibitem
\bibitem{Agostinelli:2002hh}
Geant4 collaboration, S.~Agostinelli {\em et~al.},
  \ifthenelse{\boolean{articletitles}}{\emph{{Geant4: A simulation toolkit}},
  }{}\href{https://doi.org/10.1016/S0168-9002(03)01368-8}{Nucl.\ Instrum.\
  Meth.\  \textbf{A506} (2003) 250}\relax
\mciteBstWouldAddEndPuncttrue
\mciteSetBstMidEndSepPunct{\mcitedefaultmidpunct}
{\mcitedefaultendpunct}{\mcitedefaultseppunct}\relax
\EndOfBibitem
\bibitem{LHCb-PROC-2011-006}
M.~Clemencic {\em et~al.}, \ifthenelse{\boolean{articletitles}}{\emph{{The
  \lhcb simulation application, Gauss: Design, evolution and experience}},
  }{}\href{https://doi.org/10.1088/1742-6596/331/3/032023}{J.\ Phys.\ Conf.\
  Ser.\  \textbf{331} (2011) 032023}\relax
\mciteBstWouldAddEndPuncttrue
\mciteSetBstMidEndSepPunct{\mcitedefaultmidpunct}
{\mcitedefaultendpunct}{\mcitedefaultseppunct}\relax
\EndOfBibitem
\bibitem{LHCb-DP-2018-004}
D.~M{\"u}ller, M.~Clemencic, G.~Corti, and M.~Gersabeck,
  \ifthenelse{\boolean{articletitles}}{\emph{{ReDecay: A novel approach to
  speed up the simulation at LHCb}},
  }{}\href{https://doi.org/10.1140/epjc/s10052-018-6469-6}{Eur.\ Phys.\ J.\
  \textbf{C78} (2018) 1009},
  \href{http://arxiv.org/abs/1810.10362}{{\normalfont\ttfamily
  arXiv:1810.10362}}\relax
\mciteBstWouldAddEndPuncttrue
\mciteSetBstMidEndSepPunct{\mcitedefaultmidpunct}
{\mcitedefaultendpunct}{\mcitedefaultseppunct}\relax
\EndOfBibitem
\bibitem{LHCb-PAPER-2015-031}
LHCb collaboration, R.~Aaij {\em et~al.},
  \ifthenelse{\boolean{articletitles}}{\emph{{A precise measurement of the \Bz
  meson oscillation frequency}},
  }{}\href{https://doi.org/10.1140/epjc/s10052-016-4250-2}{Eur.\ Phys.\ J.\
  \textbf{C76} (2016) 412},
  \href{http://arxiv.org/abs/1604.03475}{{\normalfont\ttfamily
  arXiv:1604.03475}}\relax
\mciteBstWouldAddEndPuncttrue
\mciteSetBstMidEndSepPunct{\mcitedefaultmidpunct}
{\mcitedefaultendpunct}{\mcitedefaultseppunct}\relax
\EndOfBibitem
\bibitem{bernlochner2021semitauonic}
F.~U. Bernlochner, M.~F. Sevilla, D.~J. Robinson, and G.~Wormser,
  \ifthenelse{\boolean{articletitles}}{\emph{{Semitauonic b-hadron decays: A
  lepton flavor universality laboratory}},
  }{}\href{https://doi.org/10.1103/RevModPhys.94.015003}{Rev.\ Mod.\ Phys.\
  \textbf{94} (2022) 015003},
  \href{http://arxiv.org/abs/2101.08326}{{\normalfont\ttfamily
  arXiv:2101.08326}}\relax
\mciteBstWouldAddEndPuncttrue
\mciteSetBstMidEndSepPunct{\mcitedefaultmidpunct}
{\mcitedefaultendpunct}{\mcitedefaultseppunct}\relax
\EndOfBibitem
\bibitem{Skwarnicki:1986xj}
T.~Skwarnicki, {\em {A study of the radiative cascade transitions between the
  Upsilon-prime and Upsilon resonances}}, PhD thesis, Institute of Nuclear
  Physics, Krakow, 1986,
  {\href{http://inspirehep.net/record/230779/}{DESY-F31-86-02}}\relax
\mciteBstWouldAddEndPuncttrue
\mciteSetBstMidEndSepPunct{\mcitedefaultmidpunct}
{\mcitedefaultendpunct}{\mcitedefaultseppunct}\relax
\EndOfBibitem
\bibitem{Caprini:1997mu}
I.~Caprini, L.~Lellouch, and M.~Neubert,
  \ifthenelse{\boolean{articletitles}}{\emph{{Dispersive bounds on the shape of
  $\Bbar\to\D^{(*)}\ell\bar{\nu}$ form factors}},
  }{}\href{https://doi.org/10.1016/S0550-3213(98)00350-2}{Nucl.\ Phys.\
  \textbf{B530} (1998) 153},
  \href{http://arxiv.org/abs/hep-ph/9712417}{{\normalfont\ttfamily
  arXiv:hep-ph/9712417}}\relax
\mciteBstWouldAddEndPuncttrue
\mciteSetBstMidEndSepPunct{\mcitedefaultmidpunct}
{\mcitedefaultendpunct}{\mcitedefaultseppunct}\relax
\EndOfBibitem
\bibitem{Boyd:1995sq}
C.~G. Boyd, B.~Grinstein, and R.~F. Lebed,
  \ifthenelse{\boolean{articletitles}}{\emph{{Model-independent determinations
  of $\Bbar\to\D\ell\bar{\nu}, \D^*\ell\bar{\nu}$ form factors}},
  }{}\href{https://doi.org/10.1016/0550-3213(95)00653-2}{Nucl.\ Phys.\
  \textbf{B461} (1996) 493},
  \href{http://arxiv.org/abs/hep-ph/9508211}{{\normalfont\ttfamily
  arXiv:hep-ph/9508211}}\relax
\mciteBstWouldAddEndPuncttrue
\mciteSetBstMidEndSepPunct{\mcitedefaultmidpunct}
{\mcitedefaultendpunct}{\mcitedefaultseppunct}\relax
\EndOfBibitem
\end{mcitethebibliography}

\newpage
\centerline
{\large\bf LHCb Collaboration}
\begin
{flushleft}
\small
R.~Aaij$^{32}$\lhcborcid{0000-0003-0533-1952},
A.S.W.~Abdelmotteleb$^{51}$\lhcborcid{0000-0001-7905-0542},
C.~Abellan~Beteta$^{45}$,
F.~Abudin{\'e}n$^{51}$\lhcborcid{0000-0002-6737-3528},
T.~Ackernley$^{55}$\lhcborcid{0000-0002-5951-3498},
B.~Adeva$^{41}$\lhcborcid{0000-0001-9756-3712},
M.~Adinolfi$^{49}$\lhcborcid{0000-0002-1326-1264},
P.~Adlarson$^{77}$\lhcborcid{0000-0001-6280-3851},
H.~Afsharnia$^{9}$,
C.~Agapopoulou$^{13}$\lhcborcid{0000-0002-2368-0147},
C.A.~Aidala$^{78}$\lhcborcid{0000-0001-9540-4988}
% ...

{\footnotesize \it

% ...
$^{9}$Universit{\'e} Clermont Auvergne, CNRS/IN2P3, LPC, Clermont-Ferrand, France\\
% ...
$^{13}$LPNHE, Sorbonne Universit{\'e}, Paris Diderot Sorbonne Paris Cit{\'e}, CNRS/IN2P3, Paris, France\\
% ...
$^{32}$Nikhef National Institute for Subatomic Physics, Amsterdam, Netherlands\\
% ...
$^{41}$Instituto Galego de F{\'\i}sica de Altas Enerx{\'\i}as (IGFAE), Universidade de Santiago de Compostela, Santiago de Compostela, Spain\\
% ...
$^{45}$Physik-Institut, Universit{\"a}t Z{\"u}rich, Z{\"u}rich, Switzerland\\
% ...
$^{49}$H.H. Wills Physics Laboratory, University of Bristol, Bristol, United Kingdom\\
% ...
$^{51}$Department of Physics, University of Warwick, Coventry, United Kingdom\\
% ...
$^{54}$School of Physics and Astronomy, University of Glasgow, Glasgow, United Kingdom\\
$^{55}$Oliver Lodge Laboratory, University of Liverpool, Liverpool, United Kingdom\\
% ...
$^{63}$Syracuse University, Syracuse, NY, United States\\
% ...
$^{77}$Department of Physics and Astronomy, Uppsala University, Uppsala, Sweden, associated to $^{54}$\\
$^{78}$University of Michigan, Ann Arbor, MI, United States, associated to $^{63}$\\
\bigskip
% ...
}
\end{flushleft}

\end{document}